\documentclass[12pt]{article}

\usepackage{tikz}
\usepackage{epsfig,latexsym,amsfonts,amsmath,amsthm,amssymb,amsbsy,multirow,slashed,wasysym,textcomp,wrapfig,datetime,comment,mathtools,cancel,cite,mathrsfs}
\usepackage[hidelinks]{hyperref}
\usetikzlibrary{calc}
\usepackage[font={footnotesize},bf]{caption}

\usepackage{subcaption}
\usepackage[font={footnotesize}]{caption}

\usepackage[normalem]{ulem} 
\numberwithin{equation}{section}
\def\ee{\end{equation}}
\def\be{\begin{equation}}
\def\bea{\begin{eqnarray}}
\def\eea{\end{eqnarray}}
\newcommand{\beq}{\begin{eqnarray}}
\newcommand{\eqq}{\end{eqnarray}}
 \newcommand{\badat}{\begin{alignedat}}
 \newcommand{\eadat}{\end{alignedat}}

\newcommand{\eal}[1]{\be \begin{aligned} #1 \end{aligned}\end{equation}} 

\newcommand{\eqn}[1]{\be #1 \end{equation}} 
\newcommand{\eqa}[1]{\bea  #1\end{eqnarray}}

\long\def\new#1\endnew{{\bf #1}}		
\long\def\del#1\enddel{}

\def\del{\partial}

\usepackage{color}

\newcommand{\pink}[1]{\textcolor{\pink}{#1}}

\definecolor{dblue}{rgb}{0.2,0.50,0.80}

\usepackage{xspace}

\newcommand{\ket}[1]{\left|#1\right\rangle}
\newcommand{\bra}[1]{\left\langle#1\right|}

\def\t2{T$^{1,1}$}

\catcode`,\active

\catcode`\,12

\usepackage{amsmath,amssymb}
\usetikzlibrary{arrows.meta,positioning,calc}
\usepackage[normalem]{ulem}

\tikzset{
  theory/.style={
    draw=black,
    line width=0.8pt,
    rounded corners=1.5pt,
    minimum height=2.35cm,
    text width=4.75cm,
    align=center,
    inner xsep=8pt,
    inner ysep=7pt,
    font=\normalsize
  },
  transition/.style={
    -{Stealth[length=3.2mm,width=2.2mm]},
    line width=0.9pt,
    black
  },
  equivalence/.style={
    {Stealth[length=3.2mm,width=2.2mm]}-{Stealth[length=3.2mm,width=2.2mm]},
    line width=0.9pt,
    black
  },
  transition label/.style={
    fill=white,
    inner sep=3pt,
    font=\normalsize,
    align=center
  }
}

\begin{document}
\begin{titlepage}
\unitlength = 1mm~\\
\vskip 3cm
\begin{center}

{\LARGE{The Stringy Photon Sphere}}

\vspace{0.8cm}
Ahmed Sheta
\vspace{1cm}

{\it  Center for the Fundamental Laws of Nature,\\ Harvard University, Cambridge, MA 02138, USA}\\

\vspace{0.5cm}
asheta@g.harvard.edu

\vspace{0.8cm}

\begin{abstract}

The S matrix on an asymptotically-flat Schwarzschild black hole is studied as a function of position on $\mathcal I$. We derive the lightcone singularities that arise in the S matrix whenever the two points on $\mathcal I$ are connected by a bulk null geodesic. This includes geodesics orbiting near the photon sphere for a long time, whose kinematics determine the associated lightcone singularities. By turning on a finite string length $\ell_s$, worldsheet corrections are shown to shift the lightcone singularities into the complex time plane, resolving the singularities at real times. In particular, we derive the leading stringy resolution of the photon sphere singularities. We also compute stringy corrections to the eikonal quasinormal mode spectrum of Schwarzschild, and discuss the kinematics of strings orbiting the photon sphere.

\end{abstract}

\end{center}

\end{titlepage}

\setcounter{tocdepth}{2}
\tableofcontents

\section{Introduction}

Studying the scattering off of near-extremal black holes \cite{Gubser:1996de,Klebanov:1997kc,Maldacena:1997ih} was a key hint in understanding their microscopic theory, which culminated with the discovery that they are described by the AdS/CFT dictionary \cite{Maldacena:1997re}. In particular, by studying the scattering problem in very extreme limits (e.g. low-energy, or more generally near the superradiant bound), a simplified analytic structure was found which hinted at the dual conformal degrees of freedom \cite{Maldacena:1997ih}.

On the other hand, there is very little handle on understanding the microscopic dynamics of asymptotically flat black holes away from extremality, Schwarzschild being the extreme example of that. In fact, there is no well-defined notion of where the microscopic degrees of freedom should live (the absorption cross-section in general does not factorize into the product of a far region term and a near-horizon term similar to \cite{Maldacena:1997ih}, which complicates the potential identification of the microscopic degrees of freedom as localized on the horizon).

Where should we look for hints about the microscopic description of asymptotically flat non-extremal black holes? A natural observable encoding the dynamics of a generic black hole is correlations of a probe field on the black hole background, the simplest correlator being the two-point function. This is not analytically tractable at arbitrary kinematics \cite{Leaver:1986gd}, since it corresponds to solving the confluent Heun differential equation \cite{Berti:2009kk,Hatsuda:2021gtn}. Inspired by the incredibly fruitful analysis of low-energy scattering on extremal black hole backgrounds, one should study kinematic limits of the scattering problem on Schwarzschild where it simplifies. For example, \cite{Dray:1984ha, tHooft:1996rdg} analyzed the correlations of particles very close to the horizon, which was later recognized to encode the chaotic behavior of black hole dynamics \cite{Polchinski:2015cea}.

A different regime in which universal behavior arises is the amplitude between two points connected by a null geodesic through the bulk, i.e. on the lightcone of each other, where the amplitude becomes singular. The lightcone singularity structure of the bulk Green's function of a massless field in the black hole background was studied in the beautiful paper \cite{Dolan:2011}, and was further analyzed in \cite{Casals:2016qyj,Buss:2017vud}, with more recent related work \cite{Grozdanov:2026cut,Grozdanov:2026ktq,Grozdanov:2026lnc}. The derivation was technically involved, relying on the detailed form of the wavefunctions deep in the bulk, which was approximated in a large-angular momentum expansion \cite{Dolan:2009nk} -- but the singularity structure at the end is a remarkably clean set of simple poles for any 2 points connected by a null geodesic. 

In this paper, we define a two-point S matrix on the black hole background, by pushing the bulk points to null infinty $\mathcal I$, and performing an LSZ-like rescaling. This S matrix can be thought of as the flat space analog of the boundary thermal two-point function in AdS. In the context of flat-space holography, this is the Carrollian two-point function \cite{Donnay:2022wvx, Alday:2024yyj, Ruzziconi:2026bix}, but studied on the nontrivial black hole background. It takes on a particularly simple form, and remains well-defined when we allow gravity to fluctuate. By a very simple WKB argument similar to \cite{Dodelson:2024bulkcone},\footnote{But a little bit simpler, since the photon sphere of a black hole in AdS is probed by eikonal \textit{imaginary}-angular momentum modes; whereas it is directly probed by modes with real angular momentum in flat space.} we can easily determine the singularities of the S matrix as a function of the kinematics on $\mathcal I$. We also provide another derivation by performing the worldline path integral to one-loop, explicitly computing the quadratic determinant. Of course, the S matrix is singular precisely when two points on $\mathcal I$ are connected by a null geodesic in the bulk; we derive the leading form of the lightcone singularities in terms of kinematics of the geodesics. We show that the singularities are the same for the three different vacua of Schwarzschild (Hartle-Hawking, Boulware, and Unruh).

\begin{figure}[h]
    \centering
    \subfloat[\centering Scattering null geodesics on the Penrose diagram. The green line denotes photon sphere geodesics, at fixed $r=3M$.] {{\includegraphics[width=0.45\textwidth]{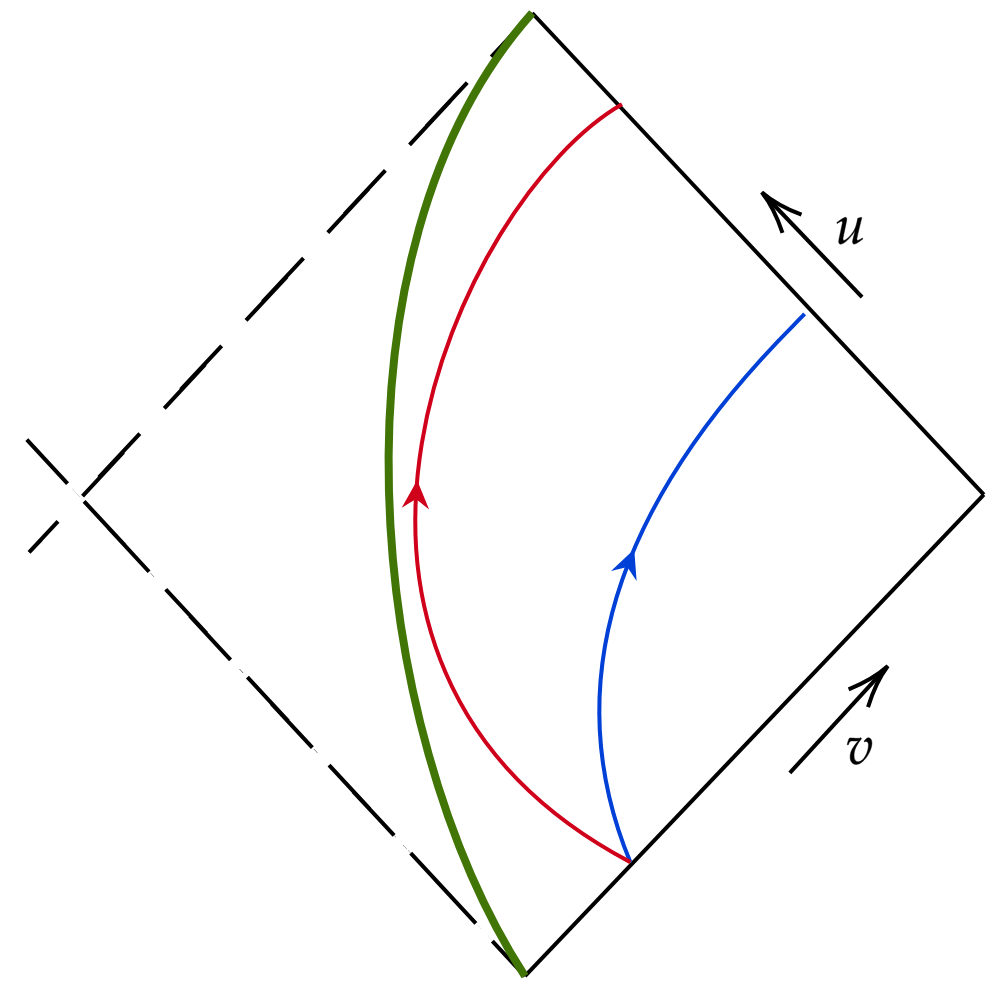} }}
    \qquad
    \subfloat[Depending on the impact parameter, geodesics could wind multiple times around the black hole before returning to $\mathcal I$. The large black circle is on $\mathcal I$, and the green circle is the photon sphere.]{{\includegraphics[width=0.45\textwidth]{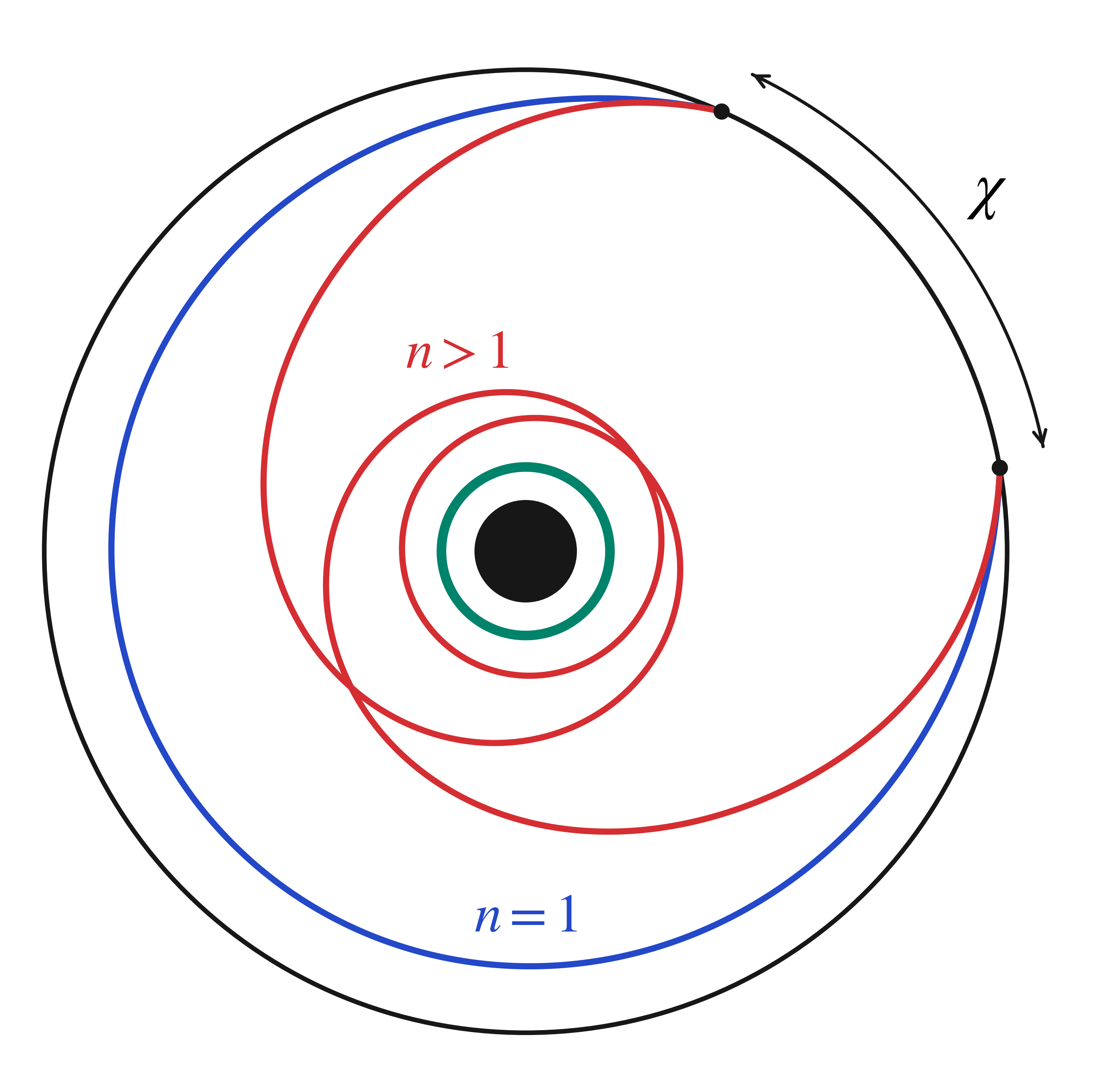} }}
    \caption{Null geodesics with $b>b_c$ do not fall into the black hole, and return to $\mathcal I^+$. The photon sphere is drawn in green. The blue geodesic has $b \gg b_c$, and gets deflected very far from the black hole, returning to $\mathcal I$ at an early time. The red geodesic has $b \to b_c$. It orbits the black hole multiple times close to the photon sphere, before returning to $\mathcal I$ at a very late time.}
    \label{fig:geodesics}
\end{figure}

In the limit of very late time on $\mathcal I^+$, the singularities probe geodesics that orbit near the photon sphere for a very long time; these are the geodesics that barely escape falling into the black hole, as shown in figure \ref{fig:geodesics}. We identify the sense in which the photon sphere is encoded in the S matrix: a tower of equally-spaced simple poles, whose spacing is controlled by the angular velocity of the photon sphere, and strength is exponentially suppressed by the photon sphere's Lyapunov exponent. In quantum gravity, these singularities would arise in the semiclassical limit whenever the local bulk description contains a photon sphere. In one of the appendices, we also provide an alternative derivation of the photon-sphere singularities using the quasinormal mode (QNM) expansion of the Green's function, by a straightforward adaptation of the derivation in \cite{Dolan:2011}, simply extrapolated to the boundary points. The singualrities are shown to be controlled by the eikonal QNM spectrum, as expected \cite{Press:1971wr,goebel_1972,schutz,Iyer:1986np,Iyer:1986nq,Cardoso:2008bp,Hadar:2022xag,Fransen:2023eqj,Kapec:2024lnr}.

The analogous singularities have been studied in AdS spacetimes. At leading order in the semiclassical limit, vacuum boundary correlators in holographic CFTs were shown to develop unexpected singularities associated with null geodesics in the empty AdS bulk \cite{Hubeny:2006yu,Gary:2009ae,Maldacena:2015iua}, which are called ``bulk-point" singularities. In the context of black hole physics, thermal CFT correlators were shown to develop singularities when the boundary points are connected by a null geodesic through the AdS-Schwarzschild bulk \cite{Dodelson:2024bulkcone}, denoted as ``bulk-cone" singularities. Of course, these singularities encode the emergent bulk locality of quantum gravity, only when quantum and stringy effects are turned off. For example, string worldsheet corrections are expected to resolve these singularities, as shown in \cite{Maldacena:2015iua} for vacuum correlators (dual to empty AdS) and \cite{Dodelson:2020lal} for thermal correlators (dual to AdS black holes). Specifically, the AdS black-hole singularities are shifted into the complex time plane, producing a finite bump on the real axis; whereas the the empty AdS singularities are completely resolved in the complex time plane. Heurestically, the null geodesic puffs up into a worldsheet of width $\ell_s$, the string length, which smoothens out the singularity. More recently, by working at finite $G_N$, quantum gravity effects were also shown to completely resolve lightcone singularities, both for empty AdS \cite{batra2026nonpromptnullraycures} and on AdS black holes \cite{AlmheiriDodelsonKomargodski2026}.

In this paper, we derive the leading stringy correction to the lightcone singularities on the asymptotically-flat Schwarzschild background. For large black holes $M \gg \ell_s$, the geodesic only probes the weak-curvature region, where perturbation theory on the string worldsheet is well-controlled. As a warmup to the string, we first provide an alternative derivation of the lightcone singularities by doing a worldline path integral in the vicinity of null geodesics, where the geometry can be approximated by a pp wave background \cite{Penrose1976,Blau:notes} and the path integral is quadratic. String propagation on pp wave backgrounds is also quadratic and exactly solveable \cite{Horowitz:1989bv,Horowitz:1990sr,Metsaev:2001bj,Berenstein:2002jq}, so we compute the worldsheet path integral similar to \cite{Dodelson:2020lal}. We show that string corrections resolve the singularities at real times. The string is sensitive to tidal effects along the null geodesic due to its finite size, which induces particle production on the worldsheet, smoothening the singularity.

More specifically, we bound the size of the S matrix for real times by $\ell_s^{-2}$. For early times, we derive a tower of singularities at imaginary times. For late times, we find a singularity in complex time, but it lies outside the validity of our approximation, so we cannot reliably predict whether the singularity has been shifted deep into the complex plane or completely resolved. However, a finite-size bump on the real axis is a prediction of our analysis. In this sense, the photon sphere is ``resolved" by string effects. In fact, we reliably compute a delay in the real-time bump relative to the expected lightcone location. Roughly speaking, due to the large accumulated tidal stretching of the string near the photon sphere, it experiences more gravitational redshift during its orbit,  delaying its arrival to $\mathcal I^+$. We interpret this in terms of a stringy correction to the eikonal QNMs, by accounting for string corrections to the photon sphere orbits.

In weakly coupled string theory ($\ell_s \gg \ell_P \sim \sqrt{G_N}$), we generically expect string effects to kick in before quantum gravity effects. Near the photon sphere, string effects are actually enhanced by the long time spent around the photon sphere, accumulating large tidal stretching. Hence, our stringy analysis is expected to be the leading correction to particle motion around the photon sphere. In this sense, the delay in the photon-sphere bump at real times is expected to persist if we turn on interactions. On the other hand, at large impact parameter, the stringy correction is suppressed due to the very weak tidal forces experienced by the string. Once we turn on interactions, we can take arbitrarily large impact parameters to reach a regime where $G_N$ effects are parametrically more important than string effects \cite{Amati:1987uf,Amati:1990xe}. It would be very interesting to apply the methods of \cite{batra2026nonpromptnullraycures} to these trajectories and see explicitly the resolution of the complex-time singularities computed in this paper. We leave this interesting question to future work. 

The singularities analyzed in this paper only probe the potential outside the photon sphere, as shown in figures \ref{fig:geodesics} and \ref{fig:WKB}. Hence, the analysis covers any geometry that looks like Schwarzschild outside the photon sphere (e.g. a single-sided black hole formed from collapse).

The outline of the paper is as follows. In section \ref{sec:kinematics}, we derive the ``S-matrix", or boundary two-point function, by extrapolating the bulk Green's function to $\mathcal I$. We show that its singularity structure is the same for all states on Schwarzchild background. In section \ref{sec:one-sided}, we show that the S matrix is singular when the two points on $\mathcal I$ are connected by a null geodesic, and derive the leading form of the singularity. We study the late-time singularities in more detail, relating it to kinematics of the photon sphere. This is reproduced by a QNM derivation in appendix \ref{app:QNM}. In section \ref{sec:worldline}, we rederive the lightcone singularities by computing the worldline path integral, matching the results of section \ref{sec:one-sided}. Section \ref{sec:string} derives the leading stringy correction of the lightcone singularities, including the imaginary-time singularities at early times and the delayed bump at late times. In section \ref{sec:stringy_qnms}, we derive string corrections to eikonal QNMs, relating them to the results of section \ref{sec:string}. We also discuss kinematics of string motion on the photon sphere. The path integral derivation on the worldline and the worldsheet are contained in appendices \ref{app:worldline} and \ref{app:worldsheet}, respectively.\footnote{After the completion of this work, I became aware of the nice paper \cite{Long:2026rpq}, which has some overlap with the analysis in sections \ref{sec:kinematics} and \ref{sec:one-sided}. I thank R. Ruzziconi for bringing this to my attention.}

\section{The Black Hole S Matrix}
\label{sec:kinematics}

We consider the four-dimensional Schwarzschild black hole,
\begin{equation}
    ds^2=-f(r)\,dt^2+\frac{dr^2}{f(r)}+r^2 d\Omega_2^2 \;,
    \qquad
    f(r)=1-\frac{2M}{r} \;.
\end{equation}
The tortoise coordinate is fixed as
\begin{equation} \label{eq:tortoise}
    r_*=r+2M\log\!\left(\frac{r}{2M}-1\right) \;,
\end{equation}
and we use the null coordinates
\begin{equation}
    u=t-r_* \;, \qquad v=t+r_* \;.
\end{equation}
Future and past null infinity, denoted by $\mathcal I^+$ and $\mathcal I^-$, are generated by $\partial_u$ and $\partial_v$ respectively.

In this section, we perform a QFT computation on a fixed background. In particular, we turn off quantum gravitational effects by setting $G_N= 0$, but we retain the background metric by keeping the Schwarzschild radius $M = G_N \cdot m$ fixed. We compute the propagator on the full curved geometry, i.e. to all-orders in $M$.  

A massless scalar field\footnote{We work with a massless scalar for simplicity in this paper, but the analysis carries through identically for gravitational perturbations. The potential \eqref{eq:RW} changes by $2M/r^3 \to 2M(1-|s|^2)/r^3$ for spin $s$ \cite{schutz}, which is irrelevant in the WKB regime we study. The spherical harmonics also change to the spin-weighted spherical harmonics \cite{Thorne,Berti:2009kk}, but they obey a similar completeness relation to \eqref{eq:completeness_spherical}, see \cite{Monteverdi:2024xyp, Chon:2003gx}. The omission of $\ell=0,1$ from the sum doesn't change the UV singularity.} has the following expansion as a quantum operator on the Schwarzschild background \cite{Christensen:1977jc,Candelas:1980zt,Buss:2017vud,Caribe:2023fhr}
\begin{equation}
    \Phi(t,r,\hat n)=\frac{1}{r}\sum_{\ell,m}\int_0^\infty \frac{d\omega}{\sqrt{4\pi\omega}}
    \sum_i\left[a^i_{\omega \ell m} e^{-i\omega t}\psi^{i}_{\omega\ell}(r) Y_{\ell m}(\hat n)+\text{h.c.}\right],
\end{equation}
where the radial wavefunction obeys the Regge--Wheeler differential equation
\begin{equation}
    \left[\frac{d^2}{dr_*^2}+\omega^2-V_\ell(r)\right]\psi_{\omega\ell}(r)=0,
    \qquad
    V_\ell(r)=f(r)\left[\frac{\ell(\ell+1)}{r^2}+\frac{2M}{r^3}\right].
\label{eq:RW}
\end{equation}
The sum $\sum_i$ is over the two sets of linearly independent modes: the in-modes
\begin{equation}
    \psi^{\rm in}_{\omega\ell}(r)\sim
    \begin{cases}
        T^{\rm in}_{\omega\ell} e^{-i\omega r_*}, & r_*\to -\infty \;, \\
        e^{-i\omega r_*}+R^{\rm in}_{\omega\ell} e^{+i\omega r_*}, & r_*\to +\infty \;.
    \end{cases}
\end{equation}
and the orthogonal up-modes 
\begin{equation}
    \psi_{\omega \ell}^{\rm up}(r) \sim \begin{cases}
        R_{\omega \ell}^{\rm up} e^{-i\omega r_*} + e^{i \omega r_*} \;, & r_* \to -\infty \;, \\
        T^{\rm up}_{\omega \ell} e^{i \omega r_*} \;, & r_* \to \infty \; \;.
    \end{cases}
\end{equation}
The natural boundary-extrapolated operator for the massless scalar at null infinity is
\begin{equation} \label{eq:bdry_operators}
    \Phi_{\mathcal I^+}(u,\hat n)=\lim_{r\to\infty} r\,\Phi(u+r_*,r,\hat n),
    \qquad
    \Phi_{\mathcal I^-}(v,\hat n)=\lim_{r\to\infty} r\,\Phi(v-r_*,r,\hat n).
\end{equation}
This operator has been studied extensively in the context of ``Carrollian holography", where it gives the Carrollian primary dual to a bulk scalar \cite{Donnay:2022wvx, Alday:2024yyj,Ruzziconi:2026bix}.

For the standard vacuum states in Schwarzschild (Boulware, Unruh, Hartle-Hawking), the bulk Feynmann function with $t>t'$ can be written as
\begin{equation} \label{eq:wightman}
    \begin{split}
        G_F^{\Omega}(u,r,\hat n;v,r',\hat n') &= -i \bra{\Omega} \Phi(u,r,\hat n) \Phi(v', r', \hat n') \ket{\Omega} \\
        &=\frac{-i}{r r'} \sum_{\ell=0}^\infty\sum_{m=-\ell}^{\ell} Y_{\ell m}(\hat n)Y_{\ell m}^*(\hat n') \\
        & \qquad \qquad \times
        \int_{-\infty}^\infty \frac{d\omega}{4\pi\omega}\,
        \Bigl[\sum_i \mathcal N_{i,\omega}^\Omega\,
        e^{-i 
        \omega r_*}\psi_{\omega \ell}^i(r) e^{-i 
        \omega r_*'} \psi^i_{\omega \ell}(r')^*\Bigr]\,
        e^{-i\omega(\mathcal T-i\epsilon)}  \;,
    \end{split}
\end{equation}
where we defined the time difference
\begin{equation} \label{eq:delta}
    \mathcal T \equiv u-v' \;.
\end{equation}
This is a subtracted version of the Shapiro time delay that remains finite on $\mathcal I$. The subtraction allows $\mathcal T$ to be negative for null geodesics, a phenoemnon studied in \cite{Penrose19801O,Cameron:2020itp}. $\mathcal N^\Omega_{i, \omega}$ is a state-dependent coefficient that measures the occupation number. We can use the relation
\begin{equation} \label{eq:completeness_spherical}
    \sum_{m=-\ell}^{\ell}Y_{\ell m}(\hat n)Y_{\ell m}^*(\hat n')
    =\frac{\ell+1/2}{2\pi}P_\ell(\cos\chi) \;,
\end{equation}
where $\cos \chi=\hat n \cdot \hat n'$, to write the propagator as
\begin{equation}
    G_F^\Omega(\mathcal T, \chi, r,r') = \frac{-i}{r r'} \sum_{\ell=0}^\infty \frac{\ell+1/2}{2\pi}P_\ell(\cos\chi) \int_{-\infty}^\infty \frac{d\omega}{4\pi\omega}\,
        \Bigl[\sum_i \mathcal N_{i,\omega}^\Omega\,
        e^{-i 
        \omega r_*}\psi_{\omega \ell,i}(r) e^{-i 
        \omega r_*'} \psi^*_{\omega \ell, i}(r')\Bigr]\,
        e^{-i\omega(\mathcal T-i\epsilon)} \;.
\end{equation}

The propagator dies off if we push the bulk points to the boundary. In particular, we take $r \to \infty$ at fixed $u=t-r_*$, and $r' \to \infty$ at fixed $v'=t'+r_*'$, sending the first point to $\mathcal I^+$ and the second point to $\mathcal I^-$. In this limit, we can rescale the propagator to get an appropriately defined S matrix
\begin{equation} \label{eq:S matrix}
    \mathcal S^\Omega(\mathcal T, \chi) \equiv \lim_{r,r' \to \infty} r r' \cdot  G_F^\Omega(\mathcal T, \chi, r,r') \;.
\end{equation}
This could be thought of as the appropriate LSZ prescription in this context, or the two-point function of the boundary operator \eqref{eq:bdry_operators}. It is very similar to the extrapolate boundary two-point function in AdS/CFT. Due to the $\partial_t \times SO(3)$ isometries of Schwarzschild, the S matrix can only depend on the time difference $\mathcal T$ and the angle between $\hat n$ and $\hat n'$, $\cos \chi = \hat n \cdot \hat n'$.\footnote{If gravity is dynamical, soft gravitons could induce an angle-dependent time shift on $\mathcal I$ \cite{Strominger:2014pwa,Hawking:2016sgy,Donnay:2018ckb}, which would render $\mathcal T$ sensitive to the BMS frame. We ignore this effect.}

To evaluate \eqref{eq:S matrix}, we can use
\begin{equation} \label{eq:lim_r_infinity}
    \sum_i \mathcal{N}^\Omega_{i,\omega} \lim_{r,r' \to \infty} e^{-i 
    \omega r_*}\psi_{\omega \ell,i}(r) e^{-i 
    \omega r_*'} \psi_{\omega \ell, i}(r')^* = \mathcal{N}^\Omega_{\rm{in}, \omega} R^{\rm in} _{\omega \ell} \;,
\end{equation}
where the other $r$-dependent terms all have the forms $e^{\pm i \omega r_*}$ or $e^{\pm i \omega r_*'}$, which away from $\omega=0$ vanish distributionally as $r_*,r_*' \to \infty$ by the Riemann-Lesbegue lemma. Subtleties with $\omega=0$ will be discussed below. Hence, we get the simple expression for the S matrix
\begin{equation} \label{eq:full_S}
    \mathcal S^\Omega(\mathcal T, \chi) = \frac{-i}{8 \pi^2} \sum_{\ell =0}^\infty (\ell + 1/2) P_{\ell}(\cos \chi) \int_{-\infty}^\infty \frac{d\omega}{\omega} \mathcal N^\Omega_{\rm in, \omega} R_{\omega \ell} \, e^{-i \omega (\mathcal T-i \epsilon)} \;,
\end{equation}
where we denote $R^{\rm in}_{\omega \ell}$ by $R_{\omega \ell}$ for the rest of this paper. This is a very natural result: the S matrix in position space on $\mathcal I$ is the Fourier transform of the reflection coefficient of an infalling mode.\footnote{This could be thought of as a Carrolian correlator, a perspective emphasized in \cite{Long:2026rpq}. Indeed, \eqref{eq:S matrix} is simply the Carrollian two-point function \cite{Alday:2024yyj, Ruzziconi:2026bix}, but on the nontrivial black hole background. The analogous object was also studied in a different basis, the Mellin basis of celestial holography, in \cite{Gonzo:2022tjm}. I thank R. Ruzziconi for discussions on this point.}

The S matrix depends on the vacuum state of the scalar field; the dependence is encoded in $\mathcal N^\Omega_{\rm{in},\omega}$. In both the Boulware and the Unruh vacua (where the in-modes are in their vacuum), we get $\mathcal N^{\rm B,U}_{\rm{in},\omega}=\theta(\omega)$, while in the Hartle--Hawking state \cite{Candelas:1980zt}
\begin{equation}
    \mathcal N^{\rm HH}_{\rm{in},\omega}=\frac{1}{1-e^{-8\pi M  \omega}} \;.
\end{equation} 

In this paper, we will focus on the singular part of the S matrix , which is given by the deep UV region of the integral ($|\omega| \to \infty$), since the integral of $R_{\omega \ell}e^{-i \omega (\mathcal T-i\epsilon)}$ over a finite region will give a finite result because $|R_{\omega \ell}| \leq 1$ \cite{Correia:2025enx}. In other words, we can isolate the singular contribution to the S matrix as
\begin{equation}
    \mathcal S_{\rm sing}^\Omega(\mathcal T, \chi) \propto \frac{-i}{8 \pi^2} \sum_{\ell =0}^\infty (\ell + 1/2) P_{\ell}(\cos \chi) \left[ \int_{\Lambda}^\infty + \int_{-\infty}^{-\Lambda}  \right] \frac{d\omega}{\omega} \mathcal N^\Omega_{\rm in, \omega} R_{\omega \ell} \, e^{-i \omega (\mathcal T-i \epsilon)} \;,
\end{equation}
for any large $\Lambda \gg M^{-1}$, with changes in $\Lambda$ giving non-singular corrections. As $|\omega| \to \infty$, all occupation numbers coincide, $\mathcal N^{\rm HH}_{\rm{in},\omega} \to \theta(\omega) = \mathcal N^{\rm B,U}_{\rm{in}, \omega}$, so the singular part is state-independent. In particular, in a thermal distribution, modes with energies much higher than the temperature are approximately in their vacuum. 

Hence, we define the singular part of the S matrix by
\begin{equation} \label{eq:singular_S}
    \mathcal S_{\rm sing}(\mathcal T, \chi) = \frac{-i}{8 \pi^2} \sum_{\ell =0}^\infty (\ell + 1/2) P_{\ell}(\cos \chi) \int_0^\infty \frac{d\omega}{\omega} R_{\omega \ell} \, e^{-i \omega (\mathcal T-i \epsilon)} \;,
\end{equation}
with the understanding that we only track the most singular dependence on $\mathcal T$, $\chi$.\footnote{There is a $\mathcal T$-independent infrared divergence from the $\omega \to 0$ end of the integral, which arises because we took the $r \to \infty$ limit \eqref{eq:lim_r_infinity} before doing the integral. This is related to a subtle soft mode that was studied in \cite{Jorstad:2024yzm}. An infrared-regulated version of \eqref{eq:full_S} was studied in \cite{Donnay:2022wvx,Long:2026rpq}. However, this is not an issue for us because \eqref{eq:singular_S} only tracks the most singular dependence on $\mathcal T$. For example, the integral could be cutoff from below at $\omega =\Lambda \gg M^{-1}$ without changing its singularity structure, with changes to $\Lambda$ giving non-singular corrections. Hence, we simply ignore this issue for the rest of the paper.
}
For real $\mathcal T$, this has the same singularity structure as the full S matrix \eqref{eq:full_S}, independent of the chosen vacuum state of the probe field. In particular, this is the full S matrix for both Unruh and Boulware vacua, but captures only the singular structure of the Hartle-Hawking vacuum. We study the singularities of \eqref{eq:singular_S} for the rest of the paper.

\section{Lightcone Singularities}
\label{sec:one-sided}

In this section, we study the singularity structure of \eqref{eq:singular_S}. We show that it is singular whenever the points on $\mathcal I^-$ and $\mathcal I^+$ are connected by a null geodesic in the bulk, and we derive the leading form of the singularity. 

\begin{figure}
    \centering
    \subfloat[\centering WKB with a first-order turning point.]{{\includegraphics[width=0.45\textwidth]{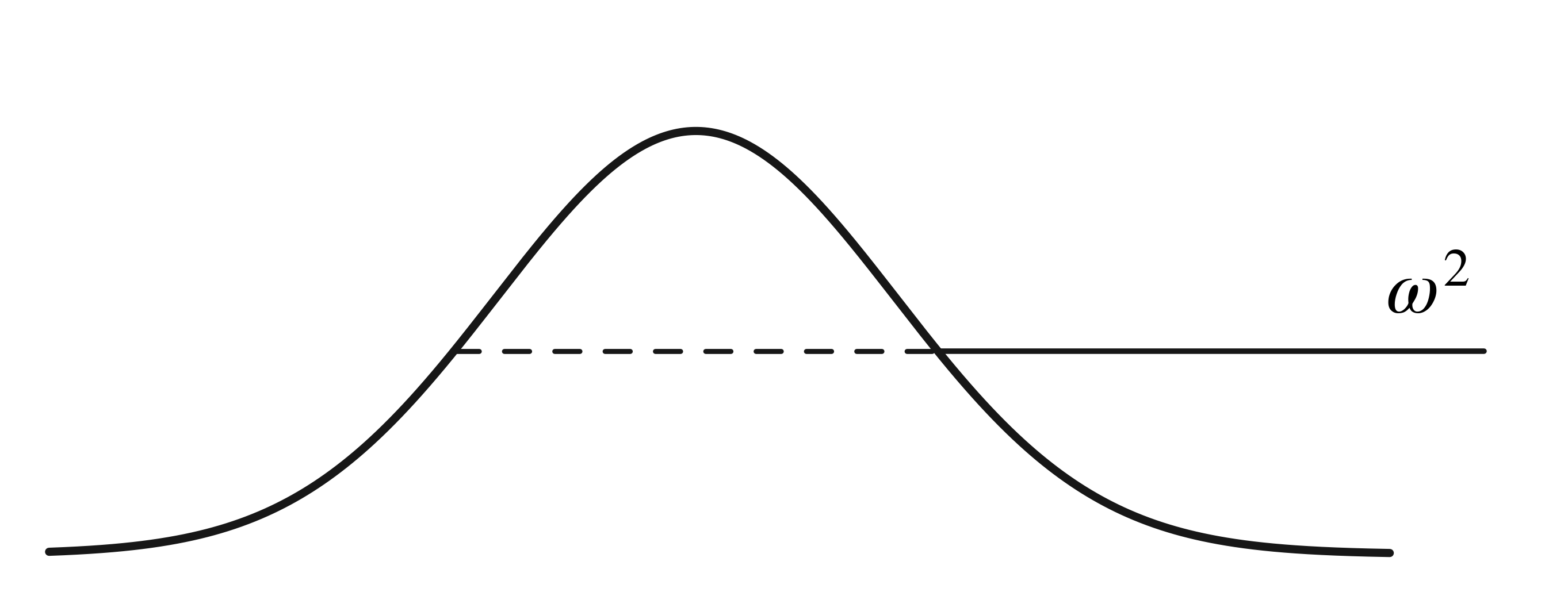} }}
    \quad
    \subfloat[WKB with a second-order turning point.]{{\includegraphics[width=0.40\textwidth]{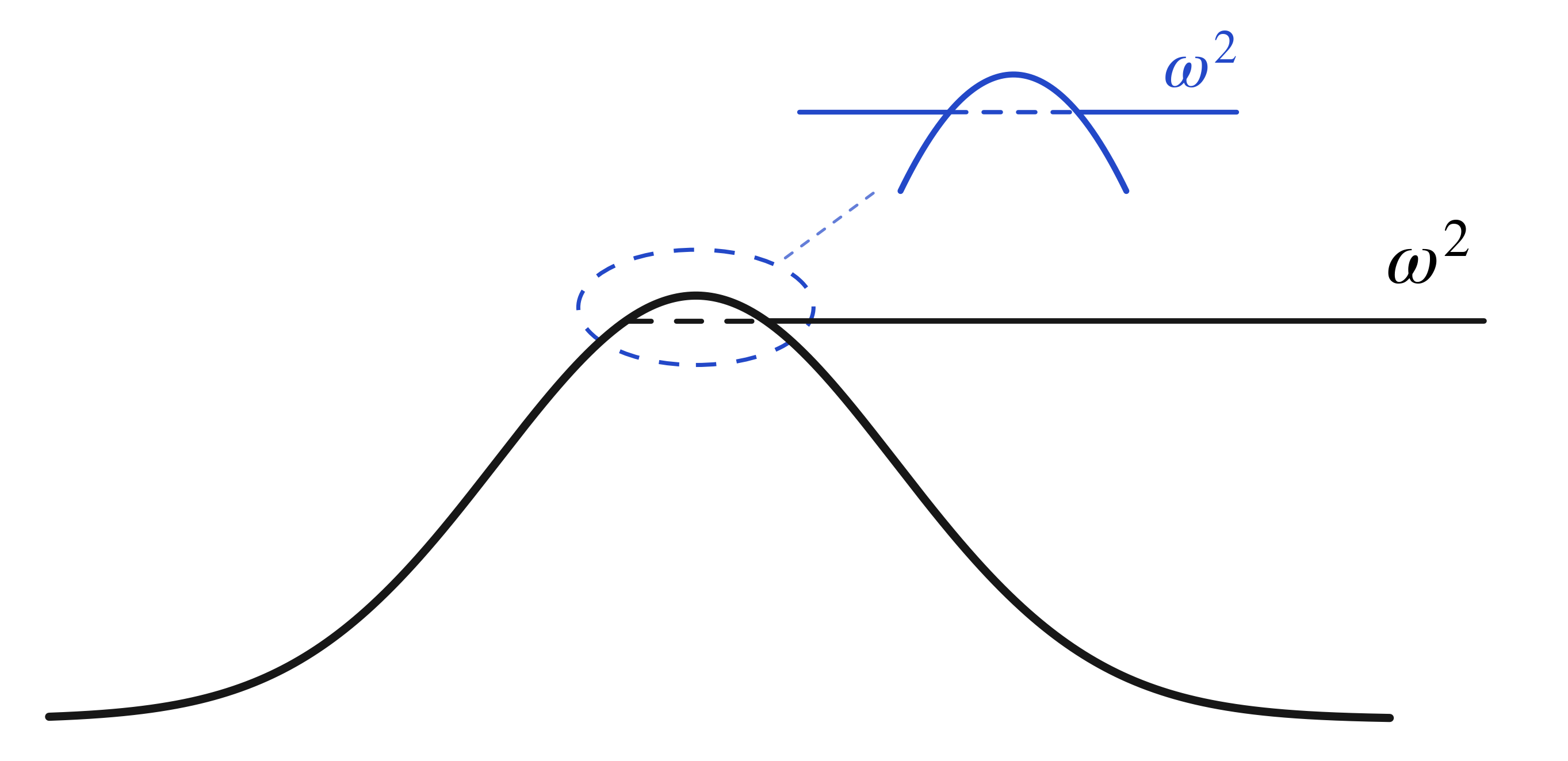} }}
    \caption{The radial wavefunction is a scattering state with the potential \eqref{eq:RW}. Depending on the ratio of $\omega/L$, we either get a first-order or second-order turning point for WKB, as $\omega \sim L \to \infty$.}
    \label{fig:WKB}
\end{figure}

As discussed in the previous section, the singularities come from the UV region of integration. At fixed $\ell$, the radial ODE \eqref{eq:RW} is a scattering problem with potential barrier $V_\ell$, depicted in figure \ref{fig:WKB}. If we scale $\omega \to \infty$ at finite $\ell$, it becomes a scattering problem where the energy is much higher than the height of the potential barrier, which gives an exponentially suppressed reflection coefficient $R_{\omega l} \sim e^{-M\omega}$. Plugging this into \eqref{eq:singular_S}, we get an analytic function of $\mathcal T$. 

Hence, to determine the singularities of \eqref{eq:singular_S}, we need to solve the radial ODE \eqref{eq:RW} for $R_{\omega \ell}$ in the limit
\begin{equation}
    \omega \gg M^{-1} \;, \qquad \ell \gg 1 \;.
\end{equation}
This is the eikonal limit where the scattering problem is controlled by geometric optics onto null geodesics, and $R_{\omega \ell}$ can be approximated by WKB. In particular, let us rewrite the kinematics as
\begin{equation}
    L \equiv \ell+1/2 \;, \qquad b\equiv L/\omega \;.
\end{equation}
The limit we are considering is then $\omega \to \infty$ at fixed $b$. In this limit, the effective radial momentum is 
\begin{equation}
    p_{r_*} \approx  \omega \sqrt{1- b^2 \frac{f(r)}{r^2}} \;.
\end{equation}

The relevant kinematic regime where $R_{\omega \ell}$ is not exponentially suppressed corresponds to scattering geodesics that connect $\mathcal I^-$ and $\mathcal I^+$. These are the geodesics that do not fall into the black hole; they have a real turning point where $p_r=0$. For a geodesic labelled by $b$, its turning point $r_t(b)$ is determined by
\begin{equation}\label{eq:turning-point}
    1- b^2 \frac{f(r_t)}{r_t^2} = 0 \;.
\end{equation}
This has a real solution for $b>b_c$, where
\begin{equation}
    b_c \equiv 3 \sqrt{3} M \;.
\end{equation}
Geodesics with $b<b_c$ simply fall into the black hole, and their corresponding wave solutions have exponentially suppressed $R_{\omega \ell}$ that does not contribute to the lightcone singularities.

We can use WKB to estimate $R_{\omega \ell}$ for $b>b_c$. Assuming a first-order turning point depicted in figure \ref{fig:WKB}a\footnote{Naively, this is only the case when $b \gg b_c$. As $b \to b_c$, we need to consider the WKB approximation with a second-order turning point \cite{schutz,Iyer:1986np}, as depicted in figure \ref{fig:WKB}b. However, as we explain in the next subsection, the first-order turning point approximation is actually sufficient for the leading form of the singularity we are considering, even for $b \approx b_c$.}, the WKB wavefunction on the right side of the barrier ($r>r_t(b)$) is given by
\begin{equation}
    \psi_\ell(r) \sim \cos \left[ \int_{r_{t*}}^{r_*} dr_*'  \sqrt{\omega^2-V_\ell(r_*')} -\frac{\pi}{4}\right] \;,
\end{equation}
where $r_{t*}$ is the tortoise coordinate corresponding to the turning poing $r_t(b)$. Separating the phase as
\begin{equation}
    \int_{r_{t*}}^{r_*} dr_*'  \sqrt{\omega^2-V_\ell(r_*')} = \int_{r_{t*}}^{r_*} dr_*'  \left( \sqrt{\omega^2-V_\ell(r_*')} - \omega \right) - \omega r_{t*} + \omega r_* \;,
\end{equation}
the reflection coefficient takes the standard form
\begin{equation} \label{eq:R_1st}
    R_{\omega \ell} \approx \exp \left[2 i \delta(\omega, L)- i \frac{\pi}{2} \right] \;,
\end{equation}
where the radial phase shift is defined as
\begin{equation} \label{eq:phase-shift}
    2 \delta (\omega, L) \equiv 2 \omega \int_{r_{t*}}^\infty dr_* \left( \sqrt{1-b^2 \frac{f(r)}{r^2}} -1 \right) - 2 \omega r_{t*} \;,
\end{equation}
which is a convergent integral, and $r$ in the integral is a function of $r_*$ defined implicitly by \eqref{eq:tortoise}. In particular, the phase $e^{2i \delta}$ comes from the propagation from $\mathcal{I}$ to the turning point, and the remaining $e^{-i\pi/2}=-i$ is the local reflection coefficient at the turning point. In flat space, we get $2 \delta (\omega , L) = -L \pi$, which gives the standard $R_{\omega \ell} = (-)^{\ell+1}$.

From a given \eqref{eq:phase-shift}, we can extract the kinematic content of the null geodesic with energy $E=\omega$ and angular momentum $L$, which play an important role in the rest of the discussion. The total angle traversed by the geodesic is given by
\begin{equation}\label{eq:deflection-angle}
    \Theta(b) =-2 \frac{\partial \delta}{\delta L} \bigg|_{\omega} = 2 \int_{r_t(b)}^\infty \frac{b\, dr}{r^2 \sqrt{1-b^2 f(r)/r^2}} \;.
\end{equation}
Note that in our conventions, $b=L/\omega>0$, so $\Theta(b)$ is always positive. It is not $2\pi$ periodic and takes winding into account.
$\Theta(b)$ is real for $b>b_c$, where it is bounded by $\Theta(b) \geq \pi$. In flat space, we get $\Theta(b) =\pi$, and the deflection angle increases in Schwarzschild due to the attractive nature of gravity. Similarly, the time difference is given by
\begin{equation} \label{eq:time-delay}
    \mathcal T (b) = 2 \frac{\partial \delta}{\partial \omega} \bigg|_{L} = 2\int_{r_t(b)}^\infty \frac{dr}{f(r)} \left[ \frac{1}{\sqrt{1-b^2 f(r)/r^2}}-1 \right] - 2 r_{t*}(b) \;.
\end{equation}
This is the precisely the boundary time difference $u_f-v_i$ of the null geodesic, which can be negative for sufficiently large $b$, as discussed in the previous section. It vanishes identically in flat space, at $M=0$.

We can use the shifted Poission resummation formula over the nonnegative integers,
\begin{equation} \label{eq:Poisson_resum}
    \sum_{\ell \geq 0} h(\ell+1/2) = \sum_{n\in \mathbb{Z}} (-)^n \int_0^\infty dL \, e^{2\pi i n L} \, h(L) \;,
\end{equation}
to rewrite the sum over $\ell$ as an $L$ integral. Moreover, in the large $\ell$ limit under consideration, we can use the Debye asymptotic approximation for $0<\chi<\pi$,
\begin{equation} \label{eq:debye}
    P_\ell(\cos \chi) \approx \sqrt{\frac{2}{\pi L \sin \chi}} \cos \left( L \chi - \frac{\pi}{4} \right) \;.
\end{equation}
The endpoints $\chi\to0$ and $\chi\to\pi$ corresponds to caustic points, where the geometric optics approximation breaks down. They can be treated by a separate approximation that is uniform in $\chi$, as studied in \cite{Dolan:2011}, but we ignore them for the rest of this paper. 

Plugging \eqref{eq:R_1st}, \eqref{eq:debye}, and \eqref{eq:Poisson_resum} into \eqref{eq:singular_S}, and changing variables of integration from $L$ to $b$, the S matrix can be written as
\begin{equation} \label{eq:S integral}
    \begin{split}
        \mathcal S_{\rm sing} (\mathcal T,\chi) &\approx \frac{-1}{16\pi^2} \cdot \sqrt{\frac{2}{\pi \sin \chi}} \sum_{n \in \mathbb Z}(-)^n \sum_{\sigma = \pm} e^{-i \sigma \pi/4} \int_0^\infty d \omega \, \omega^{1/2} \, e^{-i \omega (\mathcal T - i \epsilon)} \\
        &\qquad \qquad \qquad \qquad \qquad \qquad\times  \int_{b_c}^\infty db \, b^{1/2} \exp \{i \omega \left[W(b) + b(\sigma \chi + 2\pi n \right] \} \;.
    \end{split}
\end{equation}
We defined
\begin{equation}
    W(b) \equiv 2 \delta(\omega, L)/\omega = 2\int_{r_{t*}(b)}^\infty dr_* \left( \sqrt{1-b^2 \frac{f(r)}{r^2}}-1 \right) -2 r_{t*}(b) \;,
\end{equation}
and restored the notation $r_{t*}(b)$ to emphasize the $b$ dependence of $r_{t*}$. The $b$ integral is cutoff at $b_c$, since $b<b_c$ has exponentially suppressed contributions that cannot contribute to the singularity as explained earlier. We have only kept the leading terms, freely dropping subleading terms that do not contribute to the leading form of the singularities.

The singularities are completely determined by the $\omega \to \infty$ region of the integral, where the $b$ integral can be evaluated by a stationary phase approximation. The saddle is located at
\begin{equation} \label{eq:stationary-b}
    \frac{d}{db} \left( W(b) + b (\sigma \chi+2\pi n) \right)=0 \iff \sigma \chi+2\pi n = \Theta(b) \;.
\end{equation}
Note that if $n \notin \mathbb{Z}_+$, there will be no real solution, since $\Theta(b)>\pi$ and $0<\chi <\pi$. On the other hand, for any $n\in \mathbb{Z}_+$, $\chi$, and $\sigma$, there is a unique $b>b_c$ that solves \eqref{eq:stationary-b}. We denote this solution $b_{n,\sigma}(\chi)$, such that 
\begin{equation} \label{eq:Theta_chi}
    \Theta\left(b_{n,\sigma}(\chi)\right) = \sigma \chi + 2\pi n \;, \qquad n\in \mathbb{Z}_+ \;, \sigma=\pm \;.
\end{equation}
Note that for fixed the angular separation $\chi$ on the boundary, there are infinitely many geodesics separated by $\chi$, corresponding to all the windings ($b_{n,\sigma}$ for every $n\in \mathbb{Z}_+$ and $\sigma=\pm$), as depicted in figure \ref{fig:geodesics}b. But, of course each of these geodesics has a different time difference. Equivalently, as we derive later, at fixed $\chi$, every geodesic which solves \eqref{eq:stationary-b} contributes one term to the S matrix, but each of these terms becomes singular at a different time. 

The value of the phase at its stationary point is given by
\begin{equation} \label{eq:stationary_phase}
    W(b_{n,\sigma}(\chi)) + b_{n,\sigma}(\chi) \cdot  \Theta(b_{n,\sigma}(\chi)) = \mathcal T(b_{n,\sigma}(\chi)):= \bar{\mathcal{T}}_{n,\sigma}(\chi) \;,
\end{equation}
and the quadratic variation of the phase is given by $\omega\cdot  W''(b_{n,\sigma}(\chi)) = -\omega \cdot \Theta'(b_{n,\sigma}(\chi))$. Putting everything together, the $b$ integral in \eqref{eq:S integral} evaluates to
\begin{equation} \label{b_integral}
    \sqrt{b_{n,\sigma}(\chi)}\cdot e^{i \pi/4 \cdot  \mathrm{sgn}(W''(b))}\sqrt{\frac{2 \pi}{\omega |\Theta'(b_{n,\sigma}(\chi))|}} e^{i \omega \bar{\mathcal T}_{n,\sigma}(\chi)} \;.
\end{equation}
Using $W''(b) = -\Theta'(b) >0$ for Schwarzschild scattering with $b>b_c$, we can write down the S matrix as
\begin{equation} \label{eq:omega_integral S}
    \mathcal S_{\rm sing}(\mathcal T, \chi) \approx -\frac{1}{8 \pi^2} \cdot \sum_{n \in \mathbb{Z}_+} \sum_{ \sigma=\pm} (-)^n e^{i \pi(1-\sigma)/4} \sqrt{\frac{b_{n,\sigma}(\chi)}{\sin \chi |\Theta'(b_{n,\sigma}(\chi))|}} \int_0^\infty d\omega \, e^{-i \omega(\mathcal T-\bar{\mathcal{T}}_{n,\sigma}(\chi)-i \epsilon)} \;.
\end{equation}

The $\omega$ integral is elementary, and can be performed to give
\begin{equation} \label{eq:S final}
    \mathcal S_{\rm sing}(\mathcal T, \chi) = \frac{i}{8\pi^2} \sum_{n \in \mathbb{Z}_+} \sum_{ \sigma=\pm} \frac{\mathcal R_{n,\sigma}(\chi)}{\mathcal T-\bar{\mathcal{T}}_{n,\sigma}(\chi) -i \epsilon} \;.
\end{equation}
The residue at every pole is given by
\begin{equation} \label{eq:residue}
    \mathcal R_{n,\sigma}(\chi) = e^{-i \pi \mu_{n,\sigma}/2} \sqrt{\frac{b_{n,\sigma}(\chi)}{\sin \chi |\Theta'(b_{n,\sigma}(\chi))|}} \;,
\end{equation}
and the integer $\mu_{n,\sigma}$ is the Maslov index that counts how many caustics the geodesic has crossed \cite{Horvathy:2007nu}, and is given by
\begin{equation}
    \mu_{n,\sigma} = 2n+\frac{\sigma-1}{2} \;.
\end{equation}

Our result for the leading singular structure of the S matrix is \eqref{eq:S final}, which agrees with the large radius limit of \cite{Dolan:2011}. At fixed $\chi$, we get a discrete infinite set of singularities: whenever the time difference between the two points $\mathcal T$ coincides with  $\bar{\mathcal T}_{n,\sigma}(\chi)$ for any $n \in \mathbb{Z}_+$, $\sigma = \pm$. For each of these lightcone points, we get a simple pole with residue \eqref{eq:residue}. We ignore caustics in our discussion, so the factor of $1/\sqrt{\sin \chi}$ is finite. As the kinematics approach any given lightcone singularity $\mathcal T \to \bar{\mathcal T}_{n,\sigma}(\chi)$, the full S matrix \eqref{eq:S matrix} is approximated by the relevant term in \eqref{eq:S final}, up to corrections that are subleading for
\begin{equation}\label{eq:expansion_param}
    |\mathcal T-\bar{\mathcal T}_{n,\sigma}(\chi)| \ll M + |\Theta'(b_{n,\sigma}(\chi))|^{-1} \;.
\end{equation}
These conditions ensure that geometric optics approximation used in this section is under control, and the wave is localized on the relevant geodesic.

\subsection{Photon Sphere}
\label{sec:PR}
In this subsection, we explain in more detail how the singularity structure derived in \eqref{eq:S final} encodes the photon sphere of Schwarzschild.

The kinematics of the relevant null geodesics are given as follows. To approach the photon sphere, we take the limit
\begin{equation}
    b \to b_c \iff r_t \to 3M \;.
\end{equation}
In this limit, the traversed angle and time difference diverge logarithmically as
\begin{equation}
    \Theta(b)\simeq \mathcal L(b)+\Theta_0 \;,
    \quad
    \mathcal T (b)\simeq b_c\,\mathcal L(b)+\mathcal T_0 \;,  \qquad \text{where} ~~ \mathcal L(b) \equiv -\log\left(\frac{b-b_c}{b_c}\right) \;.
    \label{eq:photon-sphere-asymptotics}
\end{equation}
Close to the photon ring kinematics, we get the dispersion relation
\begin{equation} \label{eq:T_PR}
    \mathcal T(\Theta) \simeq b_c \Theta+ K \iff \bar{\mathcal T}_{n,\sigma}(\chi) \simeq \Omega_R^{-1}[\sigma \chi + 2\pi n] + K \;,
\end{equation}
where $\Omega_R=b_c^{-1}$ is the angular velocity at the photon sphere, and $K=k\cdot M$ for some numerical constant $k$. We also get
\begin{equation}
    \Theta'(b) \simeq -\frac{1}{b-b_c} \implies \Theta'(\mathcal T) \simeq -b_c^{-1} e^{(\mathcal T-\mathcal T_0)/b_c} =-b_c^{-1} e^{\gamma (\mathcal T-\mathcal T_0)} \;,
\end{equation}
for $\gamma = b_c^{-1}$ the Lyapunov exponent of the photon sphere.

Since $\Theta(b)$ and $\mathcal T(b)$ both diverge logarithmically, the S matrix picks up these contributions only at very late times $\mathcal T \gg M$. In particular, geodesics solving \eqref{eq:stationary-b} with $\Theta(b) \to \infty$ are precisely those with a very large winding $n \gg 1$. Hence, we can approximate the S matrix at late times by
\begin{equation}\label{eq:pr_poles_WKB}
    \begin{split}
        \mathcal S_{\rm sing}(\mathcal T, \chi) &\simeq \frac{i}{8 \pi^2} \frac{e^{\gamma \mathcal T_0/2} \,b_c}{\sqrt{\sin \chi}} \sum_{\sigma=\pm} \sum_{n \gg 1} e^{-i\pi[n+(\sigma-1)/4] } \frac{e^{-\gamma \bar{\mathcal T}_{n,\sigma}(\chi)/2}}{\mathcal T-\bar{\mathcal{T}}_{n,\sigma}(\chi)-i\epsilon} \\
        &=\frac{i}{8 \pi^2} \frac{e^{\gamma (\mathcal T_0-K)/2} \cdot b_c}{\sqrt{\sin \chi}} \sum_{\sigma=\pm} \sum_{n \gg 1} e^{-i\pi[n+(\sigma-1)/4]} \frac{e^{-(\sigma \chi + 2\pi n)/2}}{\mathcal T-\Omega_R^{-1}(\sigma \chi + 2\pi n)-K-i\epsilon}  \;.
    \end{split}
\end{equation}

This is the signature of the photon sphere in the S matrix. At fixed $\chi$ and very late time difference $\mathcal T \gg M$, we get two interlaced infinite ladders of equally-spaced singularities in $\mathcal T$, \eqref{eq:T_PR}, with successive poles separated by $\delta \mathcal T = 2\pi/\Omega_R$. The strength of these singularities is exponentially suppressed by the Lyapunov exponent \cite{Cardoso:2008bp} of the photon sphere
\begin{equation} \label{eq:PS_residues}
    |\mathcal R_{n,\sigma}(\chi)| \propto e^{-\gamma \bar{\mathcal T}_{n,\sigma}(\chi)/2} \propto e^{-(\sigma \chi+2\pi n)/2} \;.
\end{equation} 
As the scattering particle gets closer to the photon sphere ($b \to b_c$, or $\mathcal T(b) \to \infty$), a small change in $b$ corresponds to a large delay in the arrival time, $d\mathcal T \sim e^{\gamma \mathcal T(b)} |db|$. Equivalently, near-photon-sphere geodesics need to be aimed with exponentially good accuracy from $\mathcal I^-$ to reach $\mathcal I^+$ within a specified time interval.

There are two subtleties with the derivation of the photon sphere singularities that we should explain. First, we used $R_{\omega \ell} = \exp[2 i \delta - i \pi/2]$ computed in \eqref{eq:R_1st} using WKB with a first-order transition point. Close to the photon sphere, however, this assumption usually breaks down, and needs to be replaced with the inverted harmonic oscillator approximation for a second-order turning point, shown in figure \ref{fig:WKB}b \cite{schutz, Iyer:1986np, Kapec:2024lnr}. As $b \to b_c$, the appropriate WKB computation corrects the simple $-i$ local phase at the turning point into
\begin{equation} \label{eq:2nd_ord_WKB}
    R^{\rm top}_{\omega \ell} \simeq -i e^{i \beta(1-\log(\beta))} \frac{e^{\pi \beta /2}}{\sqrt{2\pi}} \Gamma(\frac{1}{2}+i \beta) \;, \qquad 
    \beta = \omega \Omega_R \frac{b^2-b_c^2}{2 b \gamma} \simeq \omega (b-b_c) \;.
\end{equation}
However, even for $b-b_c \ll b_c$, the lightcone singularity comes from the deep ultra-violet region of the $\omega$ integral, $\omega \gg (b-b_c)^{-1}$, which corresponds to a very large $\beta$. Then, we can use $R^{\rm top}_{\omega \ell} \approx -i + O(1/\beta)$. In other words, even though the first-order WKB formula at fixed $\omega$ is not \textit{uniformly} valid in $b$, the singular contributions to the S matrix come from the region $\omega \to \infty$, where \eqref{eq:R_1st} remains valid. If we compute the photon sphere S matrix using \eqref{eq:2nd_ord_WKB} instead, the computation would be valid for a bigger range than \eqref{eq:expansion_param}, but we leave this to future work.

The other subtlety concerns the use of stationary phase to evaluate the $b$ integral. At very late times, the saddle will have $b \approx b_c$. One might worry that the Gaussian integral around the saddle might be cutoff by $b_c$, the lower endpoint of the $b$ integral. However, the width of the Gaussian is given by ${ |\omega \Theta'(b)|^{-1/2} \sim [(b-b_c)/\omega]^{1/2} }$. Even though the saddle is very close to the end point of integration, the Gaussian is getting more localized for $\omega \gg (b-b_c)^{-1}$, with width much less than the distance to cutoff, $[(b-b_c)/\omega]^{1/2} \ll b -b_c$, so we safely get the full Gaussian.

\textbf{QNMs:} In appendix \ref{app:QNM}, we provide a complementary derivation of the photon sphere singularities \eqref{eq:pr_poles_WKB} from the eikonal QNM spectrum, by adapting the derivation of \cite{Dolan:2011} to the boundary propagator where it simplifies considerably, confirming the well-known equivalence between eikonal QNMs and photon sphere physics \cite{goebel_1972,schutz,Iyer:1986nq,Cardoso:2008bp,Hadar:2022xag,Fransen:2023eqj,Kapec:2024lnr}.

\section{Worldline Path Integral} \label{sec:worldline}
In this section, as a warm-up to deriving the stringy propagator, we recover the leading lightcone singularities derived in the previous section, \eqref{eq:S final}, from a worldline path integral. 

The Feynmann propagator admits the worldline path integral representation
\begin{equation} \label{eq:worldline PI}
    G_F(x_i, x_f) = -\frac{i}{2} \int_0^\infty ds \; K(x_f, x_i; s) \;,
\end{equation}
where $s$ is the modulus of the worldline (the Schwinger parameter), and
\begin{equation} \label{eq:worldline PI_K}
    K(x_i, x_f;s) = \int_{x_i}^{x_f} \mathcal D X^\mu (\lambda) \; \exp(i S[X,s]) \;, \qquad S[X,s]=\frac{1}{2s} \int_0^1 d\lambda \,g_{\mu \nu}(X) \dot X^\mu \dot X^\nu \;.
\end{equation}
The lightcone singularities are controlled by the deep UV region, corresponding to $s \to 0$ in the Schwinger parametrization. In this limit, if there is a geodesic that connects $x_i$ and $x_f$, $K(x_i, x_f;s)$ could be evaluated by the stationary phase approximation,
\begin{equation} \label{eq:g stationary phase}
    K(x_i, x_f;s) \simeq \mathcal N(s) \left( \det [\delta^2 S_\gamma(s)] \right)^{-1/2} e^{i S_\gamma(s)}  \cdot [1+\dots]\;,
\end{equation}
where $\gamma$ is the geodesic connecting $x_i$ to $x_f$, and $\mathcal{N}(s)$ is a normalization of the path integral measure. 

The singularity derived in \eqref{eq:S final} is precisely given by the phase and quadratic determinant of the contributing null geodesic \eqref{b_integral}. Hence, only the contributing geodesic and quadratic fluctuations about it are needed to reproduce a lightcone singularity from the worldline path integral. In other words, the quadratic approximation \eqref{eq:g stationary phase} is good in the regime \eqref{eq:expansion_param}.

There is a very useful scaling, the ``Penrose limit" \cite{Penrose1976}, which zooms into the geometry experienced by any null geodesic, only retaining the geodesic deviation or tidal forces around it \cite{Blau:notes}. The resulting geometry is a ``pp wave", a highly symmetric spacetime whose only nontrivial geometric content is the tidal tensor around the original geodesic. As we show, the tidal tensor exactly captures the quadratic van Vleck determinant in \eqref{eq:g stationary phase}. Hence, we should recover the leading form of any singularity in \eqref{eq:S final} by taking the Penrose limit into the contributing geodesic, and computing the worldline path integral \eqref{eq:worldline PI}-\eqref{eq:worldline PI_K} in the resulting pp wave geometry. The path integral on the pp wave is quadratic and can be performed exactly. In a sense, the pp wave ``geometrizes" WKB \cite{Fransen:2023eqj, Kapec:2024lnr}, rendering the one-loop approximation exact in the scaling limit. 

To take the Penrose limit \cite{Blau:notes}, start with a null geodesic, $x^\mu_{\rm Schw} (\tau)$, where $\tau$ is its affine parameter. It is embedded into a geodesic congruence, parametrized by 4 coordinates $(Y^-, Y^+, Y^a)$, where $Y^-$ is the affine parameter along the congruence, $Y^+$ is the Hamilton Jacobi function constructed from the momentum $dY^+=p_\mu dx_{\rm Schw}^\mu$, and $Y^a$ with $a=1,2$ are transverse coordinates. The original geodesic has $Y^+=Y^a=0$, and $Y^-=\tau$. Then, perform the rescaling $(Y^-, Y^+, Y^a) = (y^-, \varepsilon_P^2 y^+, \varepsilon_P y^a)$, and send $\varepsilon_P \to 0$ while scaling the metric as
\begin{equation} \label{eq:metric_rescaling}
    g^{\rm pp}_{\mu \nu} = \varepsilon_P^{-2} g^{\rm Schw}_{\mu \nu} \;.
\end{equation}
Finally, a coordinate transformation $(y^-, y^+, y^a) \to (x^-, x^+, x^a)$ with $x^-=y^-$ brings the pp wave metric to its Brinkmann form
\begin{equation} \label{eq:brinkmann}
    ds_{\rm pp}^2 = 2 dx^+ dx^- + A_{ab}(x^-) x^a x^b (dx^-)^2 + \delta_{ab} dx^a dx^b \;.
\end{equation}
For a null geodesic in Schwarzschild defined by $E$ and $L$, the resulting pp wave has
\begin{equation} \label{eq:A Schw}
    A_{ab}(x^-) = A(x^-) \cdot  \mathrm{diag}(1,-1) \;, \qquad A(x^-) = \frac{3 M L^2}{r(x^-)^5} \;, \qquad \dot r(x^-) = \pm \sqrt{E^2-L^2 f(r)/r^2} \;,
\end{equation}
where the dot denotes differentiation with respect to $x^-$. For the rest of this section, $x^\mu$ denotes pp wave coordiantes, and Schwarzschild coordinates will be denoted as $x_{\rm Schw}^\mu$. 

We want to perform the path integral in the pp wave geometry \eqref{eq:brinkmann}-\eqref{eq:A Schw} associated with a geodesic of fixed $E$ and $L$.\footnote{The Schwarzschild trajectory is completely fixed by $b$. $E$ labels its affine parametrization, and can be absorbed into the Schwinger parameter that is integrated over.} 
This is computed in appendix \ref{app:worldline} -- the final result \eqref{eq:G worldline final_app} is
\begin{equation} \label{eq:G worldline final}
    \lim_{r_i, r_f \to \infty} r_i r_f \cdot G_F^{\rm pp}(x_i, x_f)=  \frac{i}{8\pi^2} \frac{-E}{\Delta x^+ + \Delta(x^a\dot x^a)/2 + i\epsilon} \cdot \frac{ \sqrt{b}}{\sqrt{|\sin \Theta(b)| \cdot |\Theta'(b)|}} \cdot (-i)^{\lfloor\Theta(b)/\pi \rfloor} \;,
\end{equation}
where $\Delta x^\mu = x_f^\mu - x_i^\mu$. Here, we are focusing on a single singularity in \eqref{eq:S final}, corresponding to the $b$ geodesic we scaled onto. In particular, the two boundary points are separated by $\mathcal T \approx \mathcal T(b)$ and $\sigma \chi + 2\pi n \approx \Theta(b)$, which fixes a unique $n \in \mathbb{Z}_+$ and $\sigma\in \{\pm \}$.\footnote{\label{fn:T(b)=Tn}Alternatively, for any fixed $\chi$ and $\mathcal T$ very close to the lightcone, there is  a unique $n \in \mathbb{Z}_+$ and $\sigma=\pm$ that solves $\mathcal T \approx \bar{\mathcal T}_{n,\sigma}(\chi)$. This in turn uniquely specifies $b$ of the geodesic through the relation $\Theta(b) = \sigma \chi +2 \pi n$.} We are studying that term in \eqref{eq:S final}.

The relation between \eqref{eq:G worldline final} and the Schwarzschild propagator near the lightcone is given by
\begin{equation}
    G^{\rm Schw}_F(x_i,x_f) = \varepsilon_P^{-2} G_{F}^{\rm pp}(x_i,x_f) + \dots \;
\end{equation}
if both propagators are evaluated in the same coordinates, since the propagator solves $\Box_g G^g_F(x_1, x_2) = \delta^4(x_{12})/\sqrt{|g|}$ and the two metrics are related by \eqref{eq:metric_rescaling}. The contribution to the Schwarzschild S matrix from the pp-wave path integral is then
\begin{equation} \label{eq:S worldline}
    \mathcal S_{\rm sing} (x_i, x_f) \supset \frac{i}{8 \pi^2} \frac{1}{-\varepsilon_P^2\left[ \Delta x^+ + \Delta (x^a \dot x^a)/2 \right]/E -i \epsilon} \cdot \frac{ \sqrt{b}}{\sqrt{|\sin \Theta(b)| \cdot |\Theta'(b)|}} \cdot (-i)^{\lfloor\Theta(b)/\pi \rfloor} \;.
\end{equation}
Exactly at the endpoints of the geodesic, $\varepsilon_P^2 x^+(\pm \infty) = \varepsilon_P x^a(\pm \infty) = 0$ and \eqref{eq:S worldline} diverges. This is precisely the lightcone singularity. 

\begin{figure}[h!]
    \centering
    \includegraphics[width=7cm]{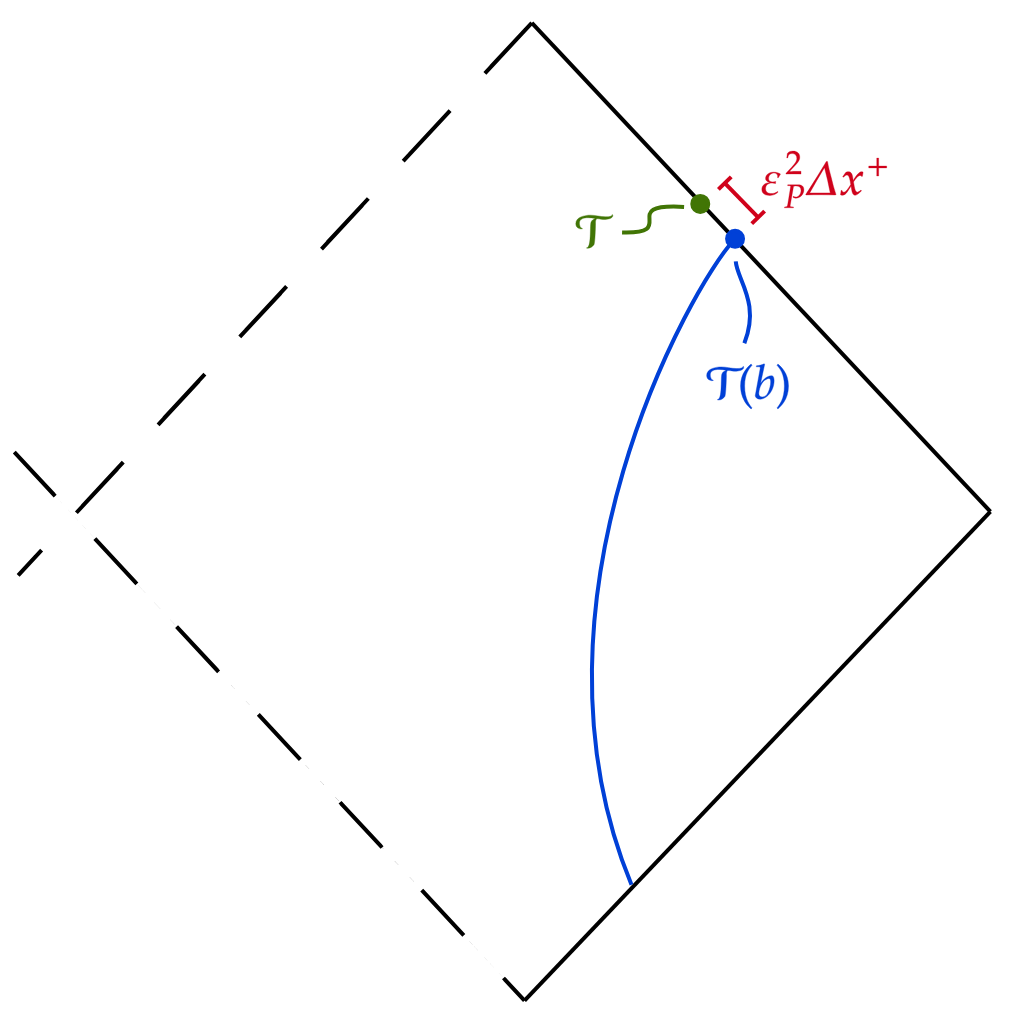}
    \caption{The blue curve depicts the reference geodesic, onto which the Penrose limit gives the pp wave background. The green mark depicts a point on $\mathcal I^+$ infinitesimally close to the endpoint of the reference geodesic. If the \textit{spatial} endpoints on $\mathcal I$ are fixed, the green point differs from the reference geodesic by its time separation, $\mathcal T -\mathcal T(b) \sim \varepsilon_P^2 \Delta x^+$, and it is not null separated from the initial point of the blue geodesic on $\mathcal I^-$.}
    \label{fig:pp_wave}
\end{figure}

However, for fixed spatial endpoints, we can study \eqref{eq:S worldline} for $\varepsilon_P^2 \Delta x^+ \neq 0$, which gives the dependence on $\mathcal T$, as shown in figure \ref{fig:pp_wave}. In this case, the Schwarzschild and pp wave coordinates are related as
\begin{equation}
    \varepsilon_P^2 \Delta x^+ = -E [\mathcal T -\mathcal T(b)] \;.
\end{equation}
Using \eqref{eq:Theta_chi}, \eqref{eq:S worldline} then simplifies to
\begin{equation}
    \mathcal S_{\rm sing}(x_i, x_f) \supset  \frac{i}{8 \pi^2}  \frac{1}{\mathcal T - \mathcal T(b)-i \epsilon} \cdot \frac{ \sqrt{b}}{\sqrt{\sin \chi \cdot |\Theta'(b)|}} \cdot e^{-i \pi [4n+\sigma-1]/4} \;,
\end{equation}
which exactly matches the $(n, \sigma)$ term in \eqref{eq:S final}.

\section{Stringy Resolution of the Lightcone} \label{sec:string}
In the previous sections, we computed lightcone singularities in the propagation of a massless scalar field on a fixed Schwarzschild background, neglecting both interactions and finite string length effects. In particular, we set both the string length $\ell_s$, and the string coupling $g_s$ (equivalently, $G_N$) to zero. Turning on either is expected to resolve lightcone singularities \cite{Maldacena:2015iua}. 

In this section, we show that stringy effects smoothen the lightcone singularities of \eqref{eq:S final}. We keep interactions suppressed $g_s=0$, and work to leading order in $\ell_s/M$, where the strings propagate on an undeformed Schwarzschild background. We follow a similar approach to \cite{Dodelson:2020lal}, expanding the worldsheet path integral around the null geodesic, and replacing the full geometry with its Penrose limit, where the path integral can be computed exactly. This is a good approximation if \begin{equation} \label{eq:expansion_param_string}
    \Theta(b) \cdot 
    \ell_s^2/M \ll M+|\Theta'(b)|^{-1} \;, \qquad |\mathcal T-{\mathcal T}(b)|\ll \ell_s^2/M \;,
\end{equation}
such that corrections to the pp-wave metric couple weakly to the worldsheet \cite{Blau:2006ar,Blau:2007db}. The second condition ensures that point-particle corrections due to deviations from the pp wave are subleading to the stringy effect we study.

The worldsheet theory consists of the 4 spacetime coordinates $X^\mu(\lambda, \sigma)$, and the remaining internal CFT, which we assume decouples from the 4d spacetime geometry.\footnote{On purely gravitational pp backgrounds, the subject of study in this section, worldsheet fermions in the superstring decouple from the background geometry \cite{Russo:2002qj,Craps:2008bv}. Hence, the fermions can be regarded as part of the internal CFT in the case of the superstring.} The spacetime scalar $\Phi$ is chosen to be a massless string state whose $X^\mu$ modes are in their ground state, which fixes the associated internal CFT state. 

Following \cite{Dodelson:2020lal}, the quantity of interest is the stringy spacetime propagator \cite{Cohen:1985sm} on the pp wave background \eqref{eq:brinkmann}. Starting from Polyakov's path integral on the cylinder, and fixing the worldsheet metric to
\begin{equation}
    ds^2_{\rm ws}=-s^2 d\lambda^2 + \ell_s^4 d\sigma^2 \;, \qquad \lambda\in[0,1] \;, ~ \sigma \in [0,2\pi] \;,
\end{equation}
the propagator is
\begin{equation} \label{eq:worldsheet PI}
\begin{split}
    & G_{\rm string}(x_i, x_f) = -\frac{i}{2} \int_0^\infty ds \, \Delta_{\rm FP}(s) \; K_{\rm int}(s) \; \mathcal N_{\rm osc} \; \int_{x_i}^{x_f} \mathcal D X^\mu \exp(i S_P[X,s]) \;, \\
    & S_P[X,s] = -\frac{1}{4\pi s} \int d\lambda \, d\sigma \; g_{\mu \nu }(X) [- \partial_\lambda X^\mu \partial_\lambda X^\nu + (s^2/\ell_s^{4}) \partial_\sigma X^\mu \partial_\sigma X^\nu] \;,
\end{split}
\end{equation}
where $s$ is the modulus of the worldsheet cylinder. 

We impose boundary conditions on the internal string modes that pick the massless spacetime scalar $\Phi$ as the initial and final string state. In particular, $X^\mu(0,\sigma)=x_i^\mu$ and $X^\mu(1,\sigma)=x_f^\mu$ fix the zero mode of $X^\mu$ at the endpoints and select a certain state for its oscillators. Upon sending the endpoints to $\mathcal I$, this state is projected to the vacuum, producing a normalization and a phase, as explained in appendix \ref{app:worldsheet}. $\mathcal N_{\rm osc}$ precisely cancels the normalization, and the phase cancels against $K_{\rm int} (s)$, the path integral over the internal CFT with appropriate boundary conditions.

The stringy propagator \eqref{eq:worldsheet PI} on the pp wave geometry \eqref{eq:brinkmann}-\eqref{eq:A Schw} is evaluated in appendix \ref{app:worldsheet}, see \eqref{eq:G_string_prop_app}.\footnote{Instead of rescaling the Schwarzschild coordinates by $\varepsilon_P$ as in \eqref{eq:metric_rescaling}, we absorp $\varepsilon_P$ into the coordinates, and regard the pp wave as the leading term in an expansion of the full metric around the geodesic \cite{Blau:2006ar}. We use this convention for the rest of the paper.} Pushing the boundary points to the boundary, and rescaling similar to \eqref{eq:S matrix}, we obtain the stringy S matrix, 
\begin{equation} \label{eq:G_string_int}
    \mathcal S_{\rm string}(\mathcal T, \chi) = \lim_{r_i, r_f} r_i r_f G_{\rm string}(x_i, x_f) = \int_0^\infty dp_+ \tilde G_0(x_i, x_f; p_+) \cdot \tilde G_{\rm osc}(x_i, x_f; p_+) \;,
\end{equation}
where $\tilde G_0(x_i, x_f, p_+)$ is the zero mode propagator (the integrand in the first line of \eqref{eq:G worldline final_app}). As discussed below \eqref{eq:G worldline final_app}, the $p_+$ integral sums over all affine parametrizations of the same trajectory in Schwarzschild. The stringy corrections are given by
\begin{equation} \label{eq:G_stringy_corr}
   \begin{split}
        \tilde G_{\rm osc}(x_i, x_f;p_+) \equiv K_{\rm int} (p_+) \cdot \prod_{a=1}^2 
        \prod_{n=1}^{\infty}  \frac{1}{2 i \sin\left(\frac{n \Delta x^-}{p_+\ell_s^2} \right) } \frac{\det[-\partial_-^2 -(n /p_+\ell_s^2)^2]}{\det \left[-\partial_-^2 + A_{aa}(\bar x^-) -(n /p_+\ell_s^2)^2 \right]} \;.
   \end{split}
\end{equation}
As derived in appendix \ref{app:worldsheet determinant}, the sin in the denominator cancels $K_{\rm int}(p_+)$, giving a very simple product over oscillator modes \eqref{eq:oscillator product}. It can be interpreted in terms of the Bogoliubov coefficients, measuring the particle production of oscillators on the worldsheet as the string traverses the Schwarzschild trajectory \cite{Dodelson:2020lal}. It approaches $1$ in the point particle limit $\ell_s \to 0$, recovering the propagator in the previous sections. 

It can be shown that \eqref{eq:G_stringy_corr} resolves the singularity coming from the integral \eqref{eq:G_string_int}. In particular, we can bound the magnitude of the S matrix as
\begin{equation}
    \begin{split}
        |\mathcal S_{\rm string}(\mathcal T, \chi) | &\leq \int_0^\infty dp_+ |\tilde G_0(x_i, x_f;p_+) | \cdot |\tilde G_{\rm osc}(x_i, x_f;p_+)| \\
        &= \frac{1}{8 \pi^2} \frac{E\sqrt{b}}{\sqrt{|\sin \Theta(b)| \cdot |\Theta'(b)|}}\int_0^\infty dp_+ |\tilde G_{\rm osc}(x_i, x_f;p_+)| \;.
    \end{split}
\end{equation}
In appendix \ref{app:bound}, we obtain the bound \eqref{eq:bound G_osc} on the magnitude of $|\tilde G_{\rm osc}(x_i, x_f;p_+)|$, which renders the integral suppressed at very large $p_+$, where the lightcone singularity originates. This bound gives
\begin{equation}
    |\mathcal S_{\rm string}(\mathcal T, \chi) | \leq \frac{1}{\ell_s^2} \cdot  \frac{1}{8 \pi^2} \frac{\sqrt{b}}{\sqrt{|\sin \Theta(b)| \cdot |\Theta'(b)|}} \left[\frac{2}{\kappa_0} + \frac{16 \kappa_0}{|\Theta'(b)|^2} \right] \;,
\end{equation}
for some number $\kappa_0$ that is fixed by the trajectory of the particle, $\kappa_0(b,M)$. The lightcone singularity in string theory is bounded by $\ell_s^{-2}$, since UV modes above $\omega \sim (\kappa_0 \ell_s^2)^{-1}$ are smoothed out. This bound is valid within the regime \eqref{eq:expansion_param_string}, where the pp wave is a good approximation.


In fact, we can do a little better. In the remainder of this section, we estimate the form of the string propagator for early times ($b_0\gg M$) and late times ($b_0 \to b_c$), as depicted in figure \ref{fig:geodesics}.

\subsection{Large Impact Parameter}
For large impact parameter $b\gg M$, the oscillator determinant can be computed, using the Born approximation for low $p_+$ and the shockwave approximation for large $p_+$ \cite{Dodelson:2020lal}. This is carried out in detail in appendix \ref{app:worldsheet determinant}, giving
\begin{equation}
    \tilde G_{\rm osc}(x_i, x_f;p_+)\simeq \frac{2\pi p_+ \ell_s^2 M E/b^2}{\sinh(2 \pi p_+ \ell_s^2 ME/b^2)} \overset{p_+ \to \infty}{\sim} 4\pi p_+ \ell_s^2 M E/b^2 \exp\left(-2\pi p_+ \ell_s^2 M E/b^2  \right) \;,
\end{equation}
which supplies the exponential suppression at large $p_+$ that smears the singularity. This also manifestly goes to 1 when $\ell_s=0$. Performing the integral gives the stringy S matrix,
\begin{equation} \label{eq:stringy_S early}
    \mathcal S_{\rm string}(\mathcal T, \chi) \simeq -\frac{1}{\ell_s^2} \frac{b^2}{32 \pi^3 M} \mathcal R(_{n, \sigma}(\chi)) \cdot \psi_1 \left[\frac{1}{2} + \frac{ib^2}{4\pi \ell_s^2 M}(\mathcal T - \mathcal T(b))\right] \;,
\end{equation}
where $\psi_1$ is the trigamma function. Using $\psi_1(1/2)=\pi^2/2$, this has a finite bump on the lightcone, 
\begin{equation} \label{eq:stringy_S early LC}
    \mathcal S_{\rm string} \big|_{\rm LC} = -\frac{1}{\ell_s^2} \cdot \frac{\mathcal R_{n,\sigma}(\chi)  b^2}{64 \pi M} \;.
\end{equation}

Perhaps a more illuminating version of \eqref{eq:stringy_S early} can be seen by using
\begin{equation}
    \frac{1}{\sinh z}=2 \sum_{m=0}^{\infty} e^{-2(m+1/2) z} \;,
\end{equation}
or equivalently the trigamma function identity
\begin{equation}
    \psi_1(z) = \sum_{m=0}^\infty \frac{1}{(z+m)^2} \;,
\end{equation}
to write \eqref{eq:stringy_S early} as
\begin{equation} \label{eq:stringy_S early v2}
    \mathcal S_{\rm string}(\mathcal T, \chi) \simeq  \frac{\mathcal R_{n,\sigma}(\chi) M}{2\pi b^2} \sum_{m=0}^{\infty} \ell_s^2 \cdot \left[\mathcal T-\mathcal T(b) - 4i\pi \ell_s^2 M(m+1/2)/b^2 \right]^{-2} \;.
\end{equation}

Stringy effects change the simple pole on the real axis into an infinite series of double poles in the complex time plane \eqref{eq:stringy_S early v2}. Some of these poles lie within the regime of validity of the pp wave approximation
\eqref{eq:expansion_param_string}, so they are reliable predictions of this computation. This is reminiscient of the tower of outgoing wavepackets in the Venenziano amplitude studied in \cite{Seiberg:2000gc}, which corresponds to an oscillating folded string that has periodic probability pulses of escaping to $\mathcal I$. It is not clear what is the precise analogous interpretation for the imaginary times encountered here, but would be very interesting to understand in the future. The imaginary shift is proportional to the small parameter $M/b$, since trajectories with very large $b$ experience weak tidal forces. On the real axis, the lightcone singularity has been smeared into the bump \eqref{eq:stringy_S early LC} of height $b^4/(M^2\ell_s^2) \gg 1$.

\subsection{Photon Sphere}
\label{sec:stringy_PS}
At very late times, the lightcone singularities are controlled by the photon sphere. The string spends a very long time \eqref{eq:photon-sphere-asymptotics} in constant orbit. It is well-known that some of the worldsheet modes become unstable in the vicinity of the photon sphere, with exponential particle production \cite{Horowitz:1989bv, Dodelson:2020lal}. We compute this exponential factor in \eqref{eq:G_osc late_time p_large}, which occurs for $Ep_+ \geq \sqrt{3} M/\ell_s^2$. It can easily be integrated, giving
\begin{equation} \label{eq:large_p+_int}
    \left| \int_{\sqrt{3} M/\ell_s^2}^\infty dp_+ \tilde G_{\rm osc}(p_+) \tilde G_0(p_+) \right|  \lesssim  \frac{1}{|\mathcal T-\mathcal T(b) - i \frac{\pi}{4 \sqrt{3}} \frac{\ell_s^2}{M} \mathcal L(b)|} \cdot \exp(-\frac{\pi}{4} \mathcal L(b)) \;.
\end{equation}
This is exponentially suppressed near the photon sphere, where $\mathcal L(b) \to \infty$. In the regime \eqref{eq:expansion_param_string}, it is subleading to the rest of the integral, studied below, so we can drop it to leading order. The singularities seem to have shifted into very large imaginary $\mathcal T$, because of the strong accumulated tidal effects on the string. However, unlike the large $b$ case, the complex singularity of \eqref{eq:large_p+_int} lies outside the regime of validity of the pp wave approximation \eqref{eq:expansion_param_string}. Within our approximation, we can not reliably know if the singularity has been completely smeared away or simply shifted deep into the complex plane.

\begin{figure}[h!]
    \centering
    \includegraphics[width=8.5cm]{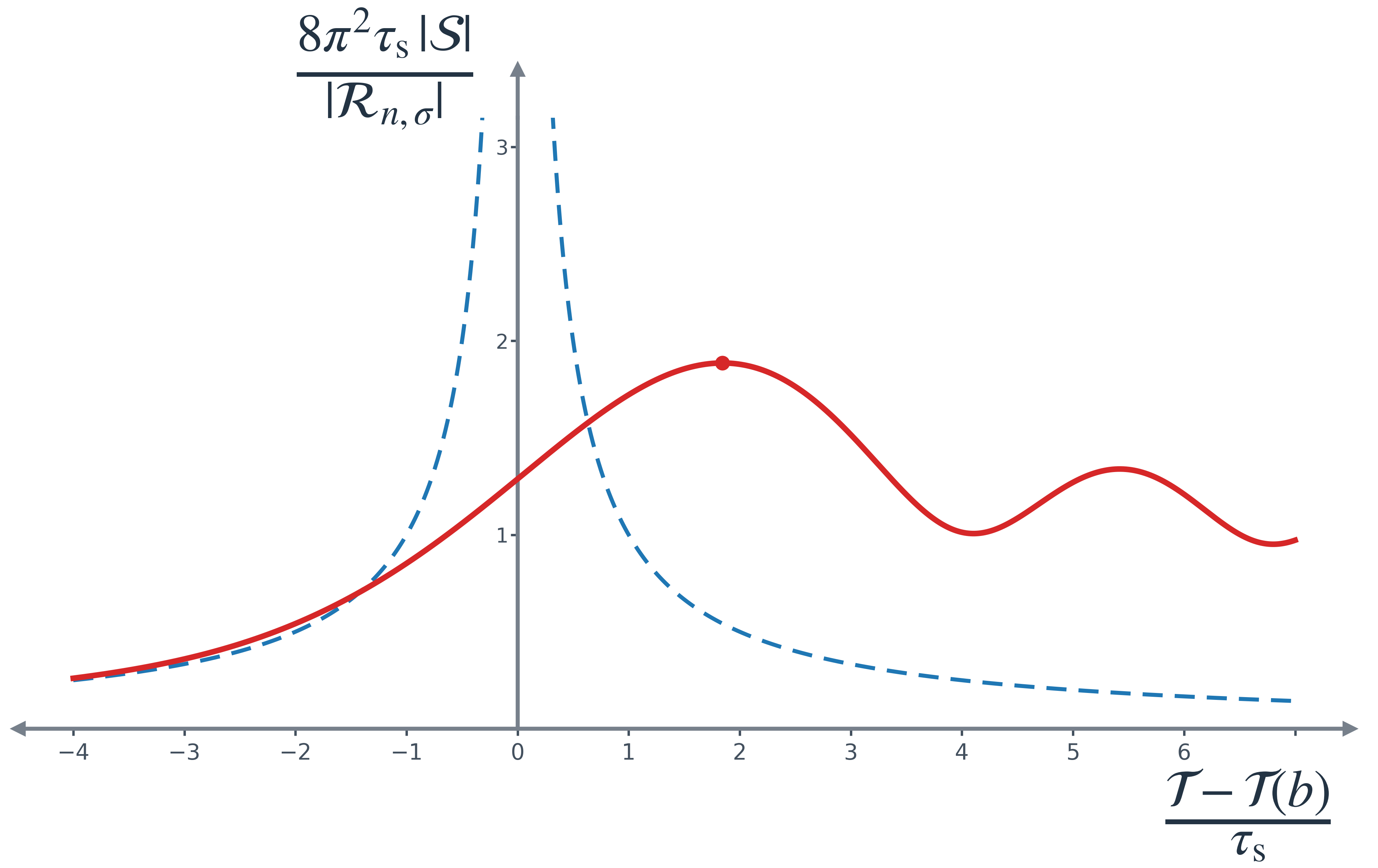}
    \caption{The magnitude of the S matrix, $|\mathcal S|$ plotted against the time on $\mathcal I^+$. The axes are normalized by $\tau_s=(\zeta(3)/4\sqrt{3}) \cdot  \ell_s^2 \mathcal L(b)/M$. The point-particle answer is plotted in dashed blue, and displays the lightcone singularity at $\mathcal T=\mathcal T(b)$. The string answer is plotted in red. It is completely finite on the real axis, and displays a delayed bump at $\mathcal T\sim \mathcal T(b)+\tau_s$.}
    \label{fig:ps_bump}
\end{figure}

The leading part of the answer is given by the rest of the integral. For $p_+$ below the cutoff, WKB is a good approximation on the worldsheet, and the oscillator determinant is approximated in \eqref{eq:G_osc late_time p_small}. Up to subleading corrections, the string S matrix is then given by
\begin{equation} \label{eq:tmp_int}
    \mathcal S_{\rm string}  \simeq \frac{-1}{8 \pi^2} \mathcal R_{n,\sigma}(\chi) \cdot \int_0^{\sqrt{3} M/\ell_s^2} d\omega \; \exp\left[-i \omega(\mathcal T- \mathcal T(b)-i \epsilon) + i \omega^3 \frac{\zeta(3)}{12 \sqrt{3}} \frac{\ell_s^6 \mathcal L(b)}{ M^3}  \right]   \;,
\end{equation}
where we define $\omega = E p_+$. The cubic term is important because $\mathcal L(b)$ is large. With a small relative error $\mathcal L(b)^{-2/3} \ll 1$, this integral can be extended to infinity, which can then be evaluated in closed form in terms of Airy and Scorer functions,
\begin{equation} \label{eq:stringy_PS}
    \mathcal S_{\rm string} \simeq -\frac{\mathcal R_{n,\sigma}(\chi)}{8 \pi} \frac{M}{\ell_s^2 \mathcal L(b)^{1/3}} \cdot c^{1/3} \left\{\mathrm{Ai}\left(-c^{1/3} \frac{M(\mathcal T-\mathcal T(b))}{\ell_s^2 \mathcal L(b)^{1/3}} \right) + i \mathrm{Gi} \left(-c^{1/3} \frac{M(\mathcal T-\mathcal T(b))}{\ell_s^2 \mathcal L(b)^{1/3}} \right) \right\} \;,
\end{equation}
where $c=4\sqrt{3}/\zeta(3)$. For example, it can be evaluated on the lightcone, giving
\begin{equation} \label{eq:bump}
    \mathcal S_{\rm string}\bigg|_{\rm LC} = \frac{-1}{8 \pi^2} \mathcal R_{n,\sigma}(\chi) e^{i\pi/6}\Gamma(4/3) \left( \frac{12\sqrt{3}}{\zeta(3)} \right)^{1/3} \frac{M}{\ell_s^2 \mathcal L(b)^{1/3}}  \;.
\end{equation}
The photon sphere's signature in the S matrix has been smeared into a finite bump. In fact, the bump is delayed from the lightcone, $\mathcal T=\mathcal T(b)$, to $\mathcal T-\mathcal T(b) \sim \ell_s^2 \mathcal L(b)^{1/3}/M$, as shown in figure \ref{fig:ps_bump}.\footnote{Technically, this lies outside the validity of the approximation \eqref{eq:expansion_param_string}. However, we can reliably compute that $\mathcal S_{\rm string}$ is increasing at $\mathcal T=\mathcal T(b)$, which indicates a delay in the bump.} Time delays due to string effects have been studied before \cite{Camanho:2014apa}. In the present context, the string gets a very large accumulated tidal stretching as it orbits the photon sphere for a long time, which slows it down and delays its arrival to $\mathcal I^+$.

In particular, near the photon sphere, null particles obey
\begin{equation}
    \frac{d\phi}{dt} \approx \frac{\sqrt{f(r)}}{r} \;,
\end{equation}
which is maximized at $r=3M$. When the string orbits near the photon sphere, $r \approx 3M$, it gets tidally stretched in the radial direction, which reduces its effective angular velocity, increasing its elapsed time. In particular, for $r<3M$, the size of the orbit is smaller, but the gravitational redshift is larger and dominates, giving a smaller angular velocity. Conversely, when $r>3M$, the gravitational redshift is weaker, but the size of the orbit is larger and dominates, also giving a smaller angular velocity. Since the string is stretched over radii $r\sim3M \pm \ell_s$, its effective angular velocity is smaller than the point particle, as depicted in figure \ref{fig:ps_pot}. We quantitatively compute this effect in the next section.

\section{Stringy Quasinormal Modes}
\label{sec:stringy_qnms}
In this section, we compute the stringy correction to the eikonal QNM spectrum of Schwarzschild, and relate it to the results of section \ref{sec:stringy_PS}. The QNMs be computed by interpreting \eqref{eq:tmp_int} as the stringy generalization of \eqref{eq:omega_integral S}, resulting from adding a stringy correction to the point-particle potential $V_{L,p}(r)=L^2 f(r)/r^2$. The stringy QNMs are then computed from the string-corrected potential using usual methods.

In particular, the phase shift \eqref{eq:phase-shift} can be written as
\begin{equation}
    \delta_p(\omega,L) = \int_{r_t}^\infty \frac{dr}{f(r)} \left( \sqrt{\omega^2-V_{L,p}(r)} - \omega \right) - \omega r_{t*} \;,
\end{equation}
where we added a subscript $p$ to indicate point-particle quantities. If we perturb the potential by a stringy correction, $V_{L,p} \to V_L = V_{L,p}+V_{L,s}$, the phase shift changes to first-order as
\begin{equation} \label{eq:phase_shift_string}
    \delta = \delta_p + \delta _s \;, \qquad \delta_s \simeq -\frac{1}{2} \int_{r_t}^\infty \frac{dr}{f(r)} \frac{V_{L,s}(r)}{\sqrt{\omega^2-V_{L,p}(r)}} \;.
\end{equation}

At the same time, using \eqref{eq:S integral} and \eqref{eq:stationary_phase}, the point-particle S matrix \eqref{eq:omega_integral S} can be written as
\begin{equation}
    \mathcal S_{\rm sing} \propto \int_0^\infty d\omega \, e^{-i \omega (\mathcal T-i\epsilon)+i  \varphi_p(\omega)} \;,
\end{equation}
where the phase is given by
\begin{equation}
    \varphi_p(\omega) = [2 \delta_p (\omega,L) + L(\sigma \chi+ 2\pi n)]_{\bar L_p} \;, \qquad 2 \partial_L \delta_p(\omega, L)\big|_{\bar L_p}+\sigma \chi+2\pi n = 0 \;.
\end{equation}
In other words, $\bar L_p$ is the saddle of the point particle integral \eqref{eq:S integral}. Of course, $\varphi_p(\omega) = \omega \mathcal T(b)$. If we change the phase-shift as \eqref{eq:phase_shift_string}, the $L$ integral would have a new saddle $\bar L=\bar L_p+ \bar L_s$, and the full string-corrected phase is given by
\begin{equation}
    \varphi(\omega) = [2\delta(\omega,L)+L(\sigma \chi+2\pi n)]_{\bar L} \;.
\end{equation}
Expanding the right-hand side to first order, writing the first-order change in the phase as $\varphi_s(\omega) = \varphi(\omega)-\varphi_p(\omega)$, and using \eqref{eq:phase_shift_string} we get
\begin{equation}
    \varphi_s(\omega) \simeq -\int_{r_t}^\infty \frac{dr}{f(r)} \frac{V_{L,s}(r)}{\sqrt{\omega^2-V_{L,p}(r)}} \;.
\end{equation}

\begin{figure}[h!]
    \centering
    \includegraphics[width=16cm]{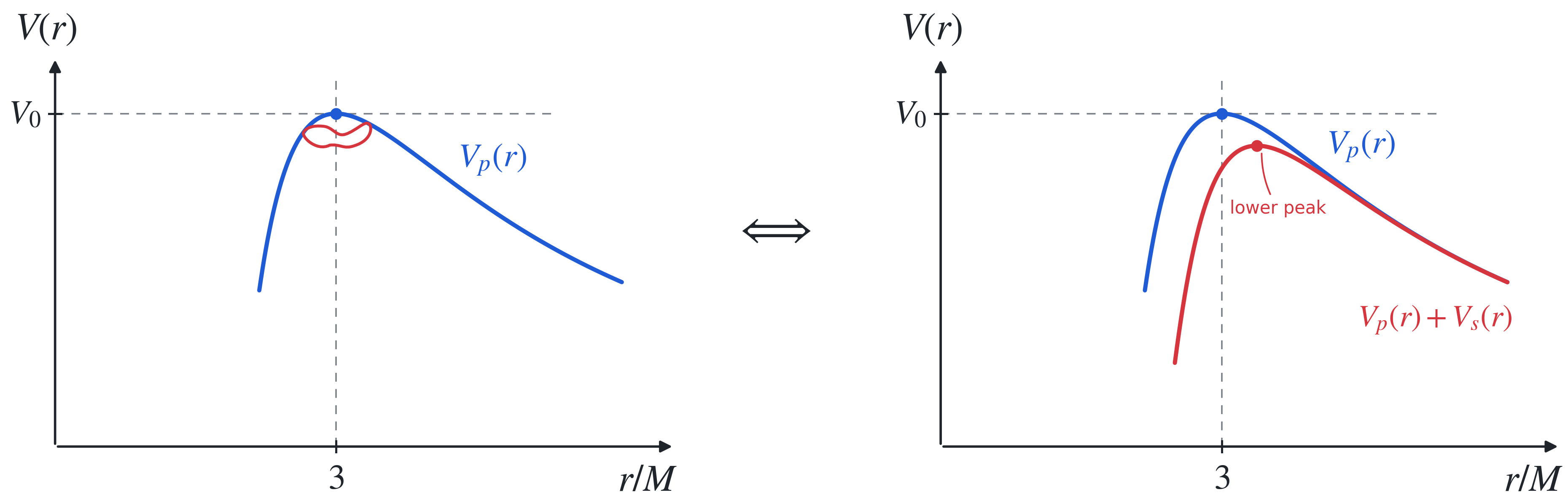}
    \caption{Averaging the string motion on the point-particle potential, as shown on the left, is equivalent to a point-object moving on the string-corrected potential, shown on the right. The peak of the potential moves downwards, giving a smaller angular velocity. The string correction is amplified for visual clarity. $V_0 =\frac{L^2}{27 M^2}$.}
    \label{fig:ps_pot}
\end{figure}

But \eqref{eq:tmp_int} gives $\varphi_s$ as the phase of $\tilde G_{\rm osc}(p_+)$, computed in \eqref{eq:G_osc_late_time_p_small_intermediate} as
\begin{equation}
    \varphi_s(\omega) = \sum_{n} \frac{(p_+ \ell_s^2)^3}{4 n^3} \int_{-\infty}^\infty dx^- A(x^-)^2 = \frac{9 \zeta(3)}{2} \ell_s^6 M^2 L^4 \int_{r_t}^\infty \frac{dr}{r^{10}\sqrt{\omega^2-V_{L,p}(r)}} \;,
\end{equation}
where we used $\omega = E p_+$, $L=\omega b$. This gives
\begin{equation}
    V_{L,s}(r) = -f(r) \frac{9 \zeta(3)}{2} \frac{\ell_s^6 M^2 L^4}{r^{10}} \;.
\end{equation}
Hence, the string-corrected string potential is given by
\begin{equation} \label{eq:full_pot}
    V_L(r) = f(r) \left[ \frac{L^2}{r^2} -\frac{9 \zeta(3)}{2} \frac{\ell_s^6 M^2 L^4}{r^{10}} \right] \;.
\end{equation}
Averaging the point-particle potential experienced by different parts of the stretched string gives the string-corrected potential, as depicted in figure \ref{fig:ps_pot}. \eqref{eq:full_pot} matches the computation in \cite{Cano:2025mht}, who considered the effects on the bulk wave equation from specific higher-derivative corrections to GR that give polarization-independent gravitational wave propagation.

This potential still has a maximum, but it is shifted outwards
\begin{equation}
    r_{\rm peak} = 3M+\frac{2 \zeta(3)}{729} \frac{\ell_s^6 L^2}{M^5} +\dots \;.
\end{equation}
Using the usual methods of WKB \cite{Iyer:1986np,Kapec:2024lnr}, the eikonal QNMs are computed as
\begin{equation} \label{eq:qnms_stringy}
    \begin{split}
        \omega_{L n} &\simeq \left[\sqrt{V_L(r_*)} - i (n+1/2) \sqrt{-V''_{L}(r_*)/ 2 V_L(r_*)} \right] \bigg|_{r_{\rm peak}} \\
        &\simeq L \Omega_R \left[1- \frac{\zeta(3)}{2916} \left( \frac{\ell_s}{M} \right)^6 L^2 \right] - i(n+1/2) \gamma \left[ 1+ \frac{5\zeta(3)}{2187} \left( \frac{\ell_s}{M} \right)^6 L^2\right] \;,
    \end{split}
\end{equation}
where $L=\ell+1/2$. This is valid in the regime where \eqref{eq:G_osc_late_time_p_small_intermediate} holds, $\omega^2 \ell_s^4/M^2 \ll 1$, or equivalently
\begin{equation}
    L \ll \frac{M^2}{\ell_s^2} \;.
\end{equation}
We also need $1 \ll L$ and $n \sim O(1)$ for the WKB approximation of the QNMs to hold. \eqref{eq:qnms_stringy} agrees with the results of \cite{Cano:2025mht}.

\subsection{Photon Sphere Kinematics of the String}
The eikonal approximation of the QNMs corresponds to the geometric optics approximation of the wave equation onto the photon sphere geodesics \cite{Cardoso:2008bp,Hadar:2022xag,Kapec:2024lnr}. In geometric optics, the dispersion relation of the wavefunction, $\omega(L)$, is related to that of the rays via $E(L)= \mathrm{Re}~ \omega(L)$. This gives the angular velocity of the rays as
\begin{equation} \label{eq:angular_velocity_dispersion}
    \frac{d\phi}{dt} = \partial_L \mathrm{Re}~\omega(L) \;.
\end{equation}
Indeed, the point-particle QNM spectrum \eqref{eq:qnms} gives $d\phi/dt=\Omega_R$, the angular velocity of null geodesics at the photon sphere. We take $\phi$ as the angle in the plane of motion.

The stringy correction to the eikonal QNMs, \eqref{eq:qnms_stringy}, could be understood geometrically in terms of the kinematics of strings on the photon sphere. For example, using \eqref{eq:angular_velocity_dispersion}, we could compute the effective angular velocity of the string on the photon sphere from \eqref{eq:qnms_stringy} as
\begin{equation} \label{eq:stringy_angular_velocity}
    \langle d\phi/dt\rangle_{\rm string} \simeq \Omega_R \left[1- \frac{\zeta(3)}{972} \left(\frac{\ell_s}{M} \right)^6 L^2\right] \;.
\end{equation}
This is a quantiative measure of the slowdown of the string relative to the point particle due to its extended size, as discussed at the end of section \ref{sec:stringy_PS}. Note that the kinematics of the string explicitly depend on $L$, equivalently $E$; the internal structure of the string causes energy-dependent gravitational scattering. \eqref{eq:stringy_angular_velocity} is valid when $L \ll M^2/\ell_s^2$.

We could turn this discussion around, and compute \eqref{eq:stringy_angular_velocity} directly from the worldsheet, then use it to compute the string correction to the eikonal QNMs. Relating Schwarzschild coordinates to pp wave coordinates near the photon sphere (see Eq 2.50 of \cite{Kapec:2024lnr}, for example) gives
\begin{equation}
    \langle d\phi/dt\rangle_{\rm string} = \Omega_R[1+(3 E^2)^{-1} \langle \partial_- X^+\rangle_{\rm string} ] \;.
\end{equation}
This can be computed by using the constraint equations on the worldsheet to express $\partial_- X^+$ in terms of the transverse coordinates $X^a$ and their worldsheet derivatives \cite{Blau:notes}, whose expectation values can be explicitly evaluated on the photon sphere pp wave. This gives
\begin{equation}
\begin{split}
    \langle d\phi/dt\rangle_{\rm string} &= \Omega_R \left[ 1- (3 E^2 p_+)^{-1} \sum_{n=1}^\infty [k_n^2 (k_n^2-A_0)^{-1/2} + k_n^2(k_n^2+A_0)^{-1/2} - 2k_n ]\right] \;,
\end{split}
\end{equation}
where $k_n$, $A_0$ are defined in \eqref{eq:kn}, \eqref{eq:A0}. Expanding in $A_0/k_n^2 \sim \omega^2 \ell_s^4/M^2 \ll 1$ and performing the sum exactly reproduces \eqref{eq:stringy_angular_velocity}. This is a direct worldsheet derivation of the stringy decrease in the photon sphere angular velocity. Plugging into \eqref{eq:angular_velocity_dispersion} can be used to compute the string correction to the real part of the QNMs, $\omega_{Ln}$, which matches the indirect computation of \eqref{eq:qnms_stringy}. 

The stringy correction to the imaginary part of $\omega_{Ln}$ could also be understood geometrically in terms of a stronger Lyapunov exponent for the string, 
\begin{equation}
    \left\langle \frac{1}{\delta r} \frac{d (\delta r)}{dt} \right\rangle _{\rm string} \simeq \gamma \left[ 1+ \frac{5\zeta(3)}{2187} \left( \frac{\ell_s}{M} \right)^6 L^2\right] \;, \qquad \delta r \equiv r-r_{\rm peak} \;.
\end{equation}

\section*{Acknowledgements}
I would like to thank Rashmish Mishra, Romain Ruzziconi, and Andy Strominger for helpful conversations. I am especially grateful to Matthew Dodelson for his many insightful questions, comments and suggestions throughout this work. I used ChatGPT for assistance with computations and checks. This work is supported by the Gordon and Betty Moore Foundation and the John Templeton Foundation via the Black
Hole Initiative, DOE grant DE-SC/0007870, and the Simons Collaboration on
Celestial Holography.

\appendix

\section{Photon Sphere Singularities from Quasinormal Modes}
\label{app:QNM}

In this appendix, we review the derivation in \cite{Dolan:2011} of the late time singularities associated with the photon sphere (with $\mathcal T \gg b_c$) from the eikonal QNMs; adapting it to the boundary extrapolated Green's function where it considerably simplifies.

The QNM expansion is easiest to see in the retarded 2-point function \cite{Leaver:1986gd,Dolan:2011},
\begin{equation}
    \begin{split}
        G^R(x,x') &= \frac{1}{4\pi^2} \frac{1}{r r'} \sum_\ell (\ell+1/2) P_\ell(\cos \chi) \int_{-\infty + i c}^{\infty+ic} d\omega e^{-i \omega(t-t')} \tilde G^R_{\omega \ell}(r,r') \\
        &= -\frac{1}{\pi} \frac{1}{rr'} \Re \bigg[ \sum_\ell (\ell+1/2)P_\ell(\cos \chi) \sum_{n=0}^\infty e^{-i \omega_{ln}(t-t')} \psi^{\rm in}_{\ell n}(r) \psi^{\rm in}_{\ell n}(r') \frac{1}{2\omega_{ln} R_{\omega \ell} T_{\omega \ell} \partial_\omega (1/T_{\omega \ell})}  \bigg] 
        \;,
    \end{split}
\end{equation}
where the contour is closed to get to the second line, and we only retain the sum over residues of $\tilde G^R$, the QNM spectrum. $\Re$ denotes the real part.

We take $r,r' \to \infty$, and define the retarded boundary propagator
\begin{equation}
    \mathcal G^R(u,\hat n; v',\hat n') = \lim_{r,r'\to \infty} r r' G^R(x,x') \;.
\end{equation}
At the QNM frequencies, the in wavefunction is purely outgoing at infinity,
\begin{equation}
    \lim_{r \to \infty} \psi^{\rm in}_{\ell n}(r)/T^{\rm in}_{\omega \ell} = (R^{\rm in}_{\omega \ell} / T_{\omega \ell}^{\rm in}) e^{i \omega_{\ell n} r_*}
\end{equation}
giving
\begin{equation}
    \mathcal G^R(u,\hat n;v',\hat n') = -\frac{1}{2\pi} \Re \bigg[ \sum_\ell (\ell+1/2)P_\ell(\cos \chi) \sum_{n=0}^\infty e^{-i \omega_{\ell n}\mathcal T} \frac{R^{\rm in}_{\omega \ell}/T^{\rm in}_{\omega \ell}}{\omega_{\ell n} \partial_\omega (1/T^{\rm in}_{\omega \ell})}  \bigg] 
    \;,
\end{equation}
where $\mathcal T=u-v'$. 

The lightcone singularities are expected to be controlled by the eikonal QNMs, $\omega, L \to \infty$. The eikonal QNMs are approximated as \cite{Iyer:1986nq,Kapec:2024lnr}
\begin{equation} \label{eq:qnms}
    \omega_{\ell n} 
    \simeq  L \Omega_R-i (n+1/2) \gamma \;, \qquad \Omega_R=\gamma=b_c^{-1} \;,
\end{equation}
giving
\begin{equation}
    \begin{split}
        \mathcal G^R(\mathcal T,\chi) &\simeq -\frac{1}{2\pi} \Re \sum_\ell(\ell+1/2)  \, P_\ell(\cos \chi)\, e^{-i L \Omega_R \mathcal T} \, \sum_n e^{-(n+1/2) \gamma\mathcal T}\frac{R^{\rm in}_{\omega \ell}/T^{\rm in}_{\omega \ell}}{ \omega_{\ell n} \partial_\omega (1/T^{\rm in}_{\omega \ell})} \\
        &\equiv -\frac{1}{2\pi} \Re \sum_\ell (\ell+1/2)  \, P_\ell(\cos \chi)\, e^{-i L \Omega_R \mathcal T}  \mathcal{F}_\ell(\mathcal T) \;.
    \end{split}
\end{equation}

The QNM excitation factors have been computed in the large $L$ limit \cite{Dolan:2011}
\begin{equation} \label{eq:excitation factors}
    \frac{R^{\rm in}_{\omega \ell}/T^{\rm in}_{\omega \ell}}{2 \omega_{\ell n} \partial_\omega (1/T^{\rm in}_{\omega \ell})} \approx L^{-1/2} e^{2i \zeta L/\sqrt{27}} \frac{(-i \kappa L)^n}{n!} i^{1/2} C \;,
\end{equation}
with $\zeta$, $C$ and $\kappa$ real constants defined in \cite{Dolan:2011}. Hence,\footnote{We perform the overtone sum explicitly. The approximation \eqref{eq:excitation factors} of the excitation factors is not uniform in $n$; it is valid at fixed $n \ll L$. Similar to \cite{Dolan:2011}, we ignore subleading terms in $n/L$. Since we correctly recover the late-time singularities of section \ref{sec:PR}, this suggests that the approximation is good for $\gamma^{-1} \ll \mathcal T$.}
\begin{equation}
\begin{split}
    \mathcal{F}_\ell(\mathcal T) &\simeq 2i^{1/2} C\,L^{-1/2}e^{2 i \zeta L/\sqrt{27}}
    \sum_{n=0}^{\infty}\frac{(-i\kappa L)^n}{n!}e^{-(n+1/2)\gamma\mathcal T} \\
    &= 2i^{1/2} C\,L^{-1/2}e^{2 i \zeta L/\sqrt{27}}
    e^{-\gamma\mathcal T/2}
    \exp\!\left[-i\kappa L e^{-\gamma\mathcal T}\right] \;.
\end{split}
\end{equation}
Using the the Debye approximation \eqref{eq:debye} and the Poisson resummation formula \eqref{eq:Poisson_resum},
\begin{equation}
    \begin{split}
        \mathcal G^R(\mathcal T, \chi) &= 
        -\frac{\sqrt{2}C}{\sqrt{\pi^3}} \frac{e^{-\gamma \mathcal T/2}}{\sqrt{\sin \chi}} \Re \; i^{1/2} \sum_{n \in \mathbb{Z}} (-)^n \int_0^\infty dL e^{2\pi i n L} e^{-i L [-2\zeta/\sqrt{27} + \Omega_R \mathcal T + \kappa e^{-\gamma \mathcal T}]} \frac{1}{2} \left( e^{i L \chi-i \pi/4 } + e^{-i L \chi + i \pi/4} \right) \\
        &\simeq -\frac{C}{\sqrt{2 \pi^3} \Omega_R} \frac{e^{-\gamma \mathcal T/2}}{\sqrt{\sin \chi}} \Re \; \sum_{n \in \mathbb{Z}} (-)^n \left(\frac{-i}{\mathcal T-b_c(\chi+2\pi n) - K - i \epsilon} + \frac{1}{\mathcal T-b_c(-\chi+2\pi n) - K - i \epsilon} \right) \;.
    \end{split}
\end{equation}
With the convention $G^R(x,x') = -i \theta(t-t') \langle \Omega | [\Phi(x), \Phi(x')]|\Omega\rangle$, the retarded correlator is related to the Feynmann propagator as
\begin{equation}
    G^R(x,x') \propto  \theta(t-t') \Re \, G^F(x,x') \;,
\end{equation}
which matches the relation between the singularities of the retarded correlator derived in this appendix and the poles of the Feynmann correlator in \eqref{eq:pr_poles_WKB}.

\section{Worldline Path Integral}
\label{app:worldline}

In this appendix, we evaluate the worldline path integral \eqref{eq:worldline PI}-\eqref{eq:worldline PI_K} in the pp wave geometry of a Schwarzschild null geodesic with energy $E$ and angular momentum $L$, \eqref{eq:brinkmann}-\eqref{eq:A Schw}.

In the pp wave, the action \eqref{eq:worldline PI_K} is given by
\begin{equation}
    S[X,s] = \frac{1}{2s}\int_0^1 d\lambda [2 \partial_\lambda X^- \partial_\lambda X^++A_{ab}[X^-] X^a X^b \partial_\lambda X^- \partial_\lambda X^- + \partial_\lambda X^a \partial_\lambda X^a] \;.
\end{equation}
Since $X^+$ appears linearly, we can integrate it out exactly
\begin{equation}
\begin{split}
    \int [DX^+] \exp\left(\frac{i}{s} X^+\partial_\lambda X^-\big|_0^1-\frac{i}{s} \int _0^1 d\lambda X^+ \partial_\lambda^2X^- \right) &= \frac{1}{2\pi s} \exp \left(\frac{i}{s} X^+\partial_\lambda X^- \big|_0^1 \right) \delta[\partial_\lambda^2 X^-]\\
    &=\frac{1}{2\pi s} \exp(i \Delta x^- \Delta x^+/s)\delta[X^- -\bar x^-] \;,
\end{split}
\end{equation}
where we used $\det[\partial_\lambda^2]=1$ on $\lambda \in[0,1]$ by the Gelfand-Yaglom theorem, and we defined the geodesic path of $x^-$,
\begin{equation} \label{eq:xm_bar}
    \bar x^-(\lambda)=x^-_i+\Delta x^- \cdot \lambda \;, \qquad \Delta x^- \equiv x^-_f-x^-_i \;.
\end{equation}
The factor $(2 \pi s)^{-1}$ comes from appropriately normalizing the measure. Then, the $X^-$ path integral is completely saturated by the delta functional
\begin{equation}
    \int DX^- \delta[X^- - \bar x^-]
    \exp\left(
        \frac{i}{2s} \int_0^1 d\lambda\,
        A_{ab}[X^-] X^a X^b
        \partial_\lambda X^- \partial_\lambda X^-
    \right) =
    \exp\left(
        \frac{i(\Delta x^-)^2}{2s} \int_0^1 d\lambda\,
        A_{ab}[\bar x^-] X^a X^b
    \right).
\end{equation}
Finally, the integral over $X^a$ is Gaussian and can be performed, giving
\begin{equation}
    \begin{aligned}
        &\int DX^a \exp\left(
            \frac{i}{2s} X^a\partial_\lambda X^a\big|_0^1
            + \frac{i}{2s} \int_0^1 d\lambda\,
            X^a[-\partial_\lambda^2+(\Delta x^-)^2A_{aa}] X^a
        \right) \\
        &\qquad \qquad \qquad \qquad  =
        \mathcal{N}_s^2
        \left(
            \frac{\det[-\partial_-^2]}
            {\det[-\partial_-^2+A_{aa}[\bar x^-]]}
        \right)^{1/2}
        \exp\left(
            \frac{i\Delta x^-}{2s}
            (x^a\dot x^a)\big|_0^1
        \right) \;,
    \end{aligned}
\end{equation}
where the dot denotes differentiation with respect to $x^-$. Here, $\mathcal{N}_s$ is an $s$-dependent normalization factor in the path integral measure,
which is fixed by comparing to the flat space propagator
\begin{equation}
    \mathcal{N}_s = \frac{e^{-i \pi/4}}{\sqrt{2\pi s}} \;.
\end{equation}

Putting everything together and restoring the appropriate $i\epsilon$ for $\Delta x^->0$, the propagator is
\begin{equation}
    \begin{aligned}
        G^{\rm pp}_F(x_i,x_f)
        &= \frac{-1}{8 \pi^2}
        \int_0^\infty \frac{ds}{s^2}
        \exp\left(
            \frac{i\Delta x^-}{s}
            \left[
                \Delta x^+ + \frac{\Delta(x^a\dot x^a)}{2}
                + i\epsilon
            \right]
        \right)
        \cdot
        \frac{\det[-\partial_{-}^2]}
        {\sqrt{
            \det[-\partial_{-}^2+A(x^-)]
            \det[-\partial_{-}^2-A(x^-)]
        }} \\
        &=\frac{-1}{8 \pi^2}
        \int_0^\infty \frac{dp_+}{\Delta x^-}
        \exp\left(
            ip_+
            \left[
                \Delta x^+ + \frac{\Delta(x^a\dot x^a)}{2}
                + i\epsilon
            \right]
        \right)
        \cdot
        \frac{\det[-\partial_{-}^2]}
        {\sqrt{
            \det[-\partial_{-}^2+A(x^-)]
            \det[-\partial_{-}^2-A(x^-)]
        }} \;.
    \end{aligned}
\end{equation}
In the last line we changed variables of integration to $p_+=\Delta x^-/s$.

The 1-loop determinant is evaluated in the next subsection, \eqref{eq:worldline 1-loop det}, which gives the pp wave propagator in the limit of $r_i, r_f \to \infty$,
\begin{equation} \label{eq:G worldline final_app}
    \begin{split}
        \lim_{r_i, r_f \to \infty} r_i r_f \cdot G_F^{\rm pp} &= \frac{-r_i r_f}{8 \pi^2}  \int_0^\infty \frac{dp_+}{\Delta x^-} \exp\left(
            ip_+
            \left[
                \Delta x^+ + \frac{\Delta(x^a\dot x^a)}{2}
                + i\epsilon
            \right]
        \right)
        \cdot \frac{\Delta x^-}{r_i r_f} \frac{E \sqrt{b} \cdot (-i)^{\lfloor\Theta(b)/\pi \rfloor}}{\sqrt{|\sin \Theta(b)| \cdot |\Theta'(b)|}} \\
        &= \frac{i}{8 \pi^2} \frac{-E}{\Delta x^+ + \Delta(x^a\dot x^a)/2 + i\epsilon} \cdot \frac{ \sqrt{b} \cdot (-i)^{\lfloor\Theta(b)/\pi \rfloor}}{\sqrt{|\sin \Theta(b)| \cdot |\Theta'(b)|}} \;.
    \end{split}
\end{equation}
In our derivation, the $p_+$ integral is roughly the integral over affine parametrizations of the same Schwarzschild trajectory. Taking the Penrose limit onto the Schwarzschild geodesic with energy $E$ fixes its affine parameter $\tau$ as $t(\tau \to \pm \infty) \sim E \tau$, and $x^-=\tau$. But the worldline integral in the pp wave geometry integrates over all of its geodesics, including those with $x^-= p_+ \tau_{\rm pp}$ for any $p_+>0$, where $\tau_{\rm pp}$ is the affine parameter in the pp wave. Lifting them to Schwarzschild, those are all the geodesics that traverse the same trajectory, but with energies $E(p_+)=p_+ \cdot  E$.

\subsection{One-loop Determinant}

We want to compute the following 2 determinants:
\begin{equation}
    D_\pm =\det[-\partial_-^2 \pm A(x^-)] \;,
\end{equation}
where
\begin{equation}
    A(x^-) = \frac{3 M L^2}{r(x^-)^5} \;, \qquad \dot r(x^-) = \pm E \sqrt{1-b^2 f(r)/r^2} \;.
\end{equation}
The functional determinant is evaluated on the space of fluctuations with Dirichlet boundary conditions, so can be computed by the Gelfand-Yaglom theorem. In particular, given a function $y_\pm(x^-)$ that solves
\begin{equation}
    [-\partial_-^2 \pm A(x^-)] y_\pm(x^-) = 0 \;,
\end{equation}
and obeys the Dirichlet boundary conditions
\begin{equation}
    y_\pm(x^-_i)=0 \;, \quad \dot y_\pm(x^-_i)=1 \;,
\end{equation}
the determinant is evaluated as
\begin{equation}
    D_\pm = y_\pm (x^-_f) \;.
\end{equation}

\subsection*{The unstable direction}
First, we solve for $D_+$. If we define the intermediate function
\begin{equation}
    f(x^-) = r(x^-)\, \dot r(x^-) \;.
\end{equation}
It solves $[-\partial_-^2+A(x^-)]f=0$. Then, we consider the function
\begin{equation} \label{eq:yp}
    y_+(x^-)= \frac{1}{E^3} f(x^-_i) f(x^-) \cdot \partial_b \Theta(b;x^-) \;,
\end{equation}
where
\begin{equation} \label{eq:Theta integrals}
    \Theta(b;x^-) = \begin{cases}
        \int_{r(x^-)}^{r_i} \frac{b\,dr}{r^2\sqrt{1-b^2 f(r)/r^2}}  & x^-<x^-_t \;,\\
        \int_{r_t(b)}^{r_i}\frac{b\,dr}{r^2\sqrt{1-b^2 f(r)/r^2}} + \int_{r_t(b)}^{r(x^-)}\frac{b\,dr}{r^2\sqrt{1-b^2 f(r)/r^2}} & x^->x^-_t \;.
    \end{cases}
\end{equation}
We can check that $[-\partial_-^2+A(x^-)]y_+=0$ and it obeys the correct boundary conditions $y_+(x^-_i)=0$, $\dot y_+(x^-_i)=1$. Hence, we get the determinant
\begin{equation} \label{eq:D+}
    D_+ = y_+(x^-_f) = r_i r_f \frac{\dot r_i \dot r_f}{E^3} \, \partial_b \Theta(b;x^-_f) \;.
\end{equation}

When we take $r_i,r_f \to \infty$, and use $\dot r_i = -\dot r_f=-E$, we get
\begin{equation} 
    \frac{1}{r_i r_f} D_+ =-\frac{1}{E} \Theta'(b) = \frac{1}{E} |\Theta'(b)| \;,
\end{equation}
where we used $\Theta'(b)<0$ for $b>b_c$. 

\subsection*{The stable direction}
For $D_-$, we have the function
\begin{equation}
    y_-(x^-) = \frac{r_i}{L} r(x^-) \cdot \sin\left[ L \cdot \int_{x^-_i}^{x^-} \frac{dx^{-'}}{r(x^{-'})^2} \right] \;.
\end{equation}
This correctly solves the ODE $[-\partial_-^2-A(x^-)]y_-(x^-)=0$ with the correct boundary conditions $y_-(x^-_i)=0, \dot y_-(x^-_i)=1$. Hence, the determinant is given by
\begin{equation}
    D_- = y_-(x^-_f)=\frac{r_i r_f}{L} \cdot \sin \left[L \cdot \int_{r_i}^{r_f} \frac{dr}{\dot r\,r^2} \right] \;,
\end{equation}
where the integral is really the sum of the infalling and outgoing pieces similar to \eqref{eq:Theta integrals}. 

Taking $r_i,r_f \to \infty$ gives
\begin{equation}
    \frac{1}{r_i r_f} D_-= \frac{1}{L} \cdot \sin \left[2 L \cdot \int_{r_t}^{\infty} \frac{dr}{\dot r\,r^2} \right] = \frac{1}{L} \cdot \sin[\Theta(b)] \;.
\end{equation}

\subsection*{Putting it together}
We should be computing the normalized determinants
\begin{equation}
    \frac{\det[-\partial_-^2\pm A(x^-)]}{\det[-\partial_-^2]} = \frac{D_\pm}{\Delta x^-} \;,
\end{equation}
where we used $\det[-\partial_-^2]=\Delta x^-$ using Gelfand-Yaglom. The normalized 1-loop determinant is given by
\begin{equation} \label{eq:worldline 1-loop det}
    \frac{\det[-\partial_-^2]}{D_+^{1/2} D_-^{1/2}} = \frac{\Delta x^-}{r_i r_f}  \cdot \frac{E\sqrt{b}}{\sqrt{|\sin\Theta(b)| \cdot |\Theta'(b)|}} (-i)^{\lfloor\Theta(b)/\pi \rfloor} \;.
\end{equation}

The phase in this equation comes from taking the square root, since $\Theta(b)>\pi$. In particular, we can write
\begin{equation}
    \sqrt{\sin \Theta(b)} = \sqrt{e^{i \pi \lfloor \Theta/\pi \rfloor} |\sin \Theta(b)|} = e^{i \pi/2 \lfloor \Theta/\pi \rfloor} \sqrt{|\sin \Theta(b)|} \;,
\end{equation}
which gives
\begin{equation}
    \frac{\det[-\partial_-^2]}{D_+^{1/2} D_-^{1/2}} =  \frac{\Delta x^-}{r_i r_f}  \cdot \frac{E\sqrt{b}}{\sqrt{|\sin\Theta(b)| \cdot |\Theta'(b)|}} \cdot (-i)^{\lfloor\Theta(b)/\pi \rfloor} \;.
\end{equation}

The phase factor could be understood more carefully as $(-i)^{N}$, where $N$ counts the number of modes with negative eigenvalue of $-\partial_-^2 - A(x^-)$ \cite{Horvathy:2007nu}. In particular, the quadratic integrals are oscillatory,
\begin{equation}
    \int_{-\infty} ^\infty df_n \, e^{i (\lambda_n+i \epsilon) f_n^2} = \sqrt{\frac{\pi}{\epsilon-i \lambda_n}} = \sqrt{\frac{\pi}{|\lambda_n|}} e^{i \pi\cdot \rm{sg}(\lambda_n)/4} \;.
\end{equation}
All the modes of the free quadratic operator $-\partial_-^2$ have positive eigenvalues. Hence, in the normalized determinant $\det[-\partial_-^2 \pm A(x^-)]/\det[-\partial_-^2]$, modes with positive eigenvalue do not contribute any phase, but each negative eigenvalue mode contribute a phase of $e^{-i \pi/4}/e^{i \pi/4} = e^{-i \pi/2} = -i$. The operator $-\partial_-^2+A(x^-)$ only has positive eigenvalues, so its normalized determinant does not contribute a phase. On the other hand, the operator $-\partial_-^2-A(x^-)$ has negative eigenvalues, whose number is the same as the number of integers satisfying $m \pi < \Theta(b)$, which is the floor function $\lfloor \Theta(b)/\pi \rfloor$. 

\section{Worldsheet Path Integral}
\label{app:worldsheet}
In this appendix, we evaluate the worldsheet path integral \eqref{eq:worldsheet PI} on the pp wave geometry of a Schwarzschild null geodesic with energy $E$ and angular momentum $L$, \eqref{eq:brinkmann}-\eqref{eq:A Schw}.
The derivation is very similar to the worldline in appendix \ref{app:worldline}.

In particular, we evaluate \eqref{eq:worldsheet PI}, where
\begin{equation}
    \begin{split}
        S_P[X,s] &=\frac1{4\pi s}\int_0^1d\lambda\int_0^{2\pi}d\sigma\,
        \Bigg[2\partial_\lambda X^-\partial_\lambda X^+
        -\frac{2s^2}{\ell_s^4}\partial_\sigma X^-\partial_\sigma X^+
        +\partial_\lambda X^a\partial_\lambda X^a
        -\frac{s^2}{\ell_s^4}\partial_\sigma X^a\partial_\sigma X^a\\
        &\qquad \qquad \qquad \qquad 
        +A_{ab}(X^-)X^aX^b
        \left((\partial_\lambda X^-)^2
        -\frac{s^2}{\ell_s^4}(\partial_\sigma X^-)^2\right)\Bigg].
    \end{split}
\end{equation}
Since $X^+$ appears linearly, it can be exactly integrated out, producing a delta functional, 
\begin{equation} \label{eq:ws_Xm_delta}
    \exp \left(i\Delta x^- \Delta x^+/s \right) \delta[(-\partial_\lambda^2 +(s^2/\ell_s^4) \partial_\sigma^2) X^- ] \;,
\end{equation}
which saturates the $X^-$ integral by fixing
\begin{equation}
    X^-(\lambda, \sigma) = \bar x^-(\lambda) \;,
\end{equation}
where $\bar x^-(\lambda)$ was defined in \eqref{eq:xm_bar}. The delta functional in \eqref{eq:ws_Xm_delta} produces the determinant factor $1/\det[-\partial_\lambda^2+(s^2/\ell_s^4) \partial_\sigma^2]$. This cancels against the Fadeev-Popov determinant
\begin{equation}
    \Delta_{\rm FP}(s) = \frac{1}{2\pi s} \det[(-\partial_\lambda^2 + (s^2/\ell_s^4) \partial_\sigma^2)] \;
\end{equation}
leaving behind $(2\pi s)^{-1}$, which captures the presence of a ghost zero mode for $b$ and for $c$ (the modulus and CKV of the cylinder). Finally, the integral over $X^a$ is Gaussian,
\begin{equation}
    \begin{split}
        &\int \mathcal DX^1 \mathcal DX^2 \exp\left( \frac{i}{4 \pi s} \int d\lambda d\sigma \left[\partial_\lambda X^a \partial_\lambda X^a - (s^2/\ell_s^4) \partial_\sigma X^a \partial_\sigma X^a + (\Delta x^-)^2  A_{ab}(\bar x^-) X^a X^b \right]  \right) \\
        &\qquad = \exp(i \Delta x^- \Delta (x^a \dot x^a)/(2s))  \\
        &\qquad \qquad \times \prod_{a=1}^2 \left\{ \mathcal N_s^0\sqrt{\frac{\det[-\partial_\lambda^2]}{\det[-\partial_\lambda^2 + (\Delta x^-)^2A_{aa}(\bar x^-)]} }
        \cdot \prod_{n=1}^{\infty}  \mathcal N_s^n \mathcal N_s^{-n} \frac{\det[-\partial_\lambda^2 -(n s/\ell_s^2)^2]}{\det \left[-\partial_\lambda^2 + (\Delta x^-)^2 A_{aa}(\bar x^-) -(n s/\ell_s^2)^2 \right]} \right\} \;.
    \end{split}
\end{equation}
The dot denotes differentiantion with respect to $x^-$, and $\mathcal N_s^n$ are all normalization factors in the path integral measure, which can be fixed by comparing to the flat space stringy propagator, giving
\begin{equation}
    \mathcal N_s^0 = \frac{e^{-i \pi/4}}{\sqrt{2\pi s}} \;, \qquad 
    \mathcal N_s^n = \mathcal N_s^{-n} = e^{-i \pi /4}\sqrt\frac{n/\ell_s^2}{2 \pi \sin(n s/\ell_s^2)} \;.
\end{equation}
As discussed in section \ref{sec:string}, we need to divide by the normalization of the vacuum wavefunction of the oscillators, to get the properly normalized $\Phi$ string state. This is a Gaussian wavefunction for each $n$ mode, evaluated at $0$, which gives
\begin{equation}
    \mathcal N_{\rm osc} = \prod_{a=1}^2 \prod_{n=1}^{\infty} \frac{\pi \ell_s^2}{n} \;.
\end{equation}

Putting everything together, the stringy propagator is given by
\begin{equation} \label{eq:G_string_prop_app}
    \begin{split}
        G_{\rm string}^{\Phi} &= -\frac{1}{8 \pi^2} \int_0^\infty \frac{dp_+}{ \Delta x^-} \, K_{\rm int}^\Phi (p_+) \,\exp \left(i p_+[\Delta x^+ + \Delta (x^a \dot x^a)/2 + i \epsilon] \right) \\
        & \qquad \times \prod_{a=1}^2 \left\{ \sqrt \frac{\det[-\partial_-^2]}{\det[-\partial_-^2 + A_{aa}(\bar x^-)]} 
        \cdot 
        \prod_{n=1}^{\infty}  \frac{1}{2 i \sin\left(\frac{n \Delta x^-}{p_+\ell_s^2} \right) } \frac{\det[-\partial_-^2 -(n /p_+\ell_s^2)^2]}{\det \left[-\partial_-^2 + A_{aa}(\bar x^-) -(n /p_+\ell_s^2)^2 \right]} \right\}\;,
    \end{split}
\end{equation}
where we changed variables of integration $p_+ = \Delta x^-/s$. This nicely separates into the worldline propagator smeared against the stringy corrections, as written in \eqref{eq:G_string_int}-\eqref{eq:G_stringy_corr}. 

In the remainder of this appendix, we obtain bounds and approximations on the oscillator determinant when the endpoints are pushed to the boundary, $r_i, r_f \to \infty$.

\subsection{One-loop Oscillator Determinant}
\label{app:worldsheet determinant}

Define
\begin{equation} \label{eq:kn}
    D_{n,\pm} = \det[-\partial_-^2\pm A(x^-) - k_n^2] \;, \qquad k_n \equiv \frac{n}{p_+ \ell_s^2} \;,
\end{equation}
where $A$ is given by \eqref{eq:A Schw}. From the boundary conditions on the path integral \eqref{eq:worldsheet PI}, the determinant is evaluated over the space of fluctuations with Dirichlet boundary conditions. Hence, it can be evaluated by the Gelfand-Yaglom theorem, with functions $y_{n, \pm}(x^-)$ subject to 
\begin{equation} \label{eq:GY ws}
    (-\partial_-^2\pm A(x^-) - k_n^2) y_{n,\pm}(x^-) = 0 \;, \qquad y_{n,\pm}(x^-_i)=0 \;, ~~ \dot y_{n,\pm}(x_i^-)=1 \;.
\end{equation}

As $x^- \to \pm \infty$, $A(x^-) \to 0$ since $r(x^-) \to \infty$. The early time behavior is then given by 
\begin{equation} \label{eq:yi}
    y_{n,\pm}(x^-) = \frac{1}{k_n} \sin \left( k_n (x^--x^-_i) \right) \;, \qquad \qquad \text{for } x^- \to -\infty \;,
\end{equation}
where the normalization was set by the boundary conditions in \eqref{eq:GY ws}. Similarly the late time behavior is given by
\begin{equation} \label{eq:yf}
    y_{n,\pm}(x^-) = A_{n, \pm} \exp \left( ik_n (x^--x^-_i) \right) + B_{n, \pm} \exp \left( -ik_n (x^--x^-_i) \right) \;, \qquad \text{for  } x^- \to +\infty \;,
\end{equation}
where $A$ and $B$ are coefficeints determined by solving the differential equation \eqref{eq:GY ws}. 

If we take $x_f^- \to +\infty$, the determinant $D_{\pm, n}=y_{\pm,n}(x_f^-)$ is then given as
\begin{equation} \label{eq:yf_D}
\begin{split}
    D_{n, \pm} & = \alpha_{n, \pm} \exp \left( ik_n \Delta x^- \right) + \beta_{n, \pm} \exp \left( -ik_n \Delta x^- \right) \\
    &\to \alpha_{n,\pm} \exp \left(ik_n \Delta x^- \right) \;,
\end{split}
\end{equation}
where the $x^-(1-i \epsilon)$ prescription was used to project to the vacuum of the oscillator wavefunction (the sign of $i\epsilon$ is fixed, since $x^-$ flows in the same direction as Schwarzschild time). \eqref{eq:yf_D} is equivalent to evaluating the determinant in terms of the Bogoliubov coefficients for particle production on the worldsheet, as understood in \cite{Dodelson:2020lal}. In particular, as $x^-_i \to-\infty(1-i \epsilon)$, the early time solution \eqref{eq:yi} approaches $e^{i k_n (x^--x^-_i)}/2ik_n$, so the coefficient $2i k_n \alpha_{n,\pm}$ is one of the Bogoliubov coefficeints, whose magnitude gives the particle production number $|2i k_n \alpha_{n,\pm}| = \sqrt{1+\langle N_{\rm prod} \rangle}$. However, we are retaining the phase of $\alpha_{n,\pm}$ here.

The determinant is normalized by the flat space determinant, computed by Gelfand-Yaglom
\begin{equation}
    \det [-\partial_-^2 - k_n^2] = \frac{1}{k_n} \sin \left( k_n \Delta x^- \right)  \;.
\end{equation}
Hence, the oscillator determinant \eqref{eq:G_stringy_corr} reduces to
\begin{equation}
    K_{\rm int}(p_+)\exp\left(-2i \Delta x^- \cdot \sum_{n=1}^\infty k_n\right) \prod_{n=1}^\infty \frac{-1}{4k_n^2} \frac{1}{\alpha_{n,+} \cdot \alpha_{n,-}} \;,
\end{equation}
and is determined by solving the differential equation \eqref{eq:GY ws} for the coefficients $\alpha_{n,\pm}$, $\beta_{n,\pm}$. The phase is, of course, the worldsheet energy of the asymptotic vacuum state of the $X^\mu$ oscillators, $\exp({-i E_{\rm osc}^0s})$. Since the internal CFT doesn't probe the Schwarzschild geometry, we simply get $K_{\rm int}(s) =\langle E_{\rm int}^\Phi| e^{-i H_{\rm int} s} |E_{\rm int}^\Phi\rangle = \exp({-i E_{\rm int}^\Phi s})$, where $E_{\rm int}^\Phi$ is the worldsheet energy of the internal CFT state that selects $\Phi$ as the string state. By the mass-shell condition, $E_{\rm osc}^0+E^\Phi_{\rm int} \propto m_\Phi^2 =0$, so $K_{\rm int}(s)$ cancels against the phase from oscillator path integral, and \eqref{eq:G_stringy_corr} is simply given by
\begin{equation} \label{eq:oscillator product}
    \lim_{r_i, r_f \to \infty} \tilde G_{\rm osc}(x_i,x_f;p_+)=\prod_{n=1}^\infty \frac{-1}{4k_n^2} \frac{1}{\alpha_{n,+}\cdot \alpha_{n,-}} \;.
\end{equation}

\subsubsection{Bound on the Magnitude}
\label{app:bound}
First, we derive a bound on $| \tilde G_{\rm osc}(x_i, x_f; p_+)|$. Since the ODE \eqref{eq:GY ws} has real coefficients, the following quantity is conserved for any solution $f(x^-)$,
\begin{equation}
    \partial_-(\dot f f^* - f \dot f^*)=0 \;.
\end{equation}
Applying this to the solution
\begin{equation}
    f(x^-) = \begin{cases}
        e^{ik_n x^-} \;, \qquad \qquad \qquad \qquad \qquad ~~~ x^- \to -\infty \;, \\
        c_n e^{i k_n x^-} + d_n e^{-i k_n x^-} \;, \qquad \qquad  x^- \to +\infty \;,
    \end{cases}
\end{equation}
we get
\begin{equation}
    |c_n|^2 -|d_n|^2=1 \implies |c_n| \geq 1 \;.
\end{equation}
But $f$ is related to our solution \eqref{eq:yi} via
\begin{equation} 
    y(x^-) = \frac{1}{2i k_n} [f^*(x_i^-) f(x^-) - f(x_i^-) f^*(x^-)] \implies \alpha_{n,\pm} = \frac{1}{2i k_n} (c_n-d_n^* e^{2i k_n x_i^-}) \to \frac{1}{2i k_n} c_n \;,
\end{equation}
where we took $x_i^- \to -\infty(1-i \epsilon)$. Hence, we get the bound
\begin{equation} \label{eq:bound1}
    2 k_n |\alpha_{n, \pm}| \geq 1 \implies 4 k_n^2 |\alpha_{n,+} \cdot \alpha_{n,-}| \geq 1 \;.
\end{equation}

We could get a slightly stronger bound using our results from the point-particle determinant. In particular, if we allow $k_n$ in the ODE \eqref{eq:GY ws} to vary continuously as $k$, then the early time solution \eqref{eq:yi} has the correct $k \to 0$ limit to match the point-particle solution for the Gelfand-Yaglom determinant,
\begin{equation}
    y_{k,\pm} (x^- \to -\infty) \xrightarrow{k \to 0} x^- - x_i^- \;.
\end{equation}
The coefficients $\alpha_{k,\pm}$ could be extracted from the late-time solution \eqref{eq:yf_D} in the limit $x_i^- \to -\infty$, $x_f^- \to \infty$,
\begin{equation}
    e^{-i k \Delta x^-} \cdot [\partial_{x^-_f} + i k][\partial_{x_i^-} - i k] y_{k,\pm}(x^-) = 4k^2 \alpha_{k, \pm} \;.
\end{equation}
Applying this to the zero-mode \eqref{eq:D+}, we get
\begin{equation}
    \lim_{k \to 0} 4 k^2 \alpha_{k,+} = \partial_{x_i^-} \partial_{x_f^-}D_+ = -E^2 \partial_{r_i} \partial_{r_f} D_+ = -E | \Theta'(b)| \implies \lim_{k \to 0} 4k^2 |\alpha_{k,+}| > \frac{E |\Theta'(b)|}{2} \;.
\end{equation}
Since \eqref{eq:GY ws} depends smoothly on $k$, $k^2\alpha_{k,+}$ also depends smoothly on $k$, and so there exists some neighborhood of $k=0$ where
\begin{equation}
    4k^2 |\alpha_{k,+}| > \frac{E |\Theta'(b)|}{2} \;.
\end{equation}
This inequality holds for $0 \leq  k<k_0$ for some $k_0$ that is fixed by the pp wave profile $A(x^-)$. In fact, from \eqref{eq:GY ws} and \eqref{eq:A Schw}, rescaling the energy of the Schwarzschild geodesic $E \to \lambda E$ corresponds to rescaling $x^- \to \lambda^{-1} x^-$ and $k \to \lambda k$. Hence, we can write $k_0 = E \cdot \kappa_0$, where $\kappa_0$ is fixed by the trajectory of the geodesic (i.e. $\kappa_0(b,M)$). 

Applying this to the $\alpha_{n,+}$, we get
\begin{equation}
    2 k_n |\alpha_{n,+}| > \frac{E |\Theta'(b)|}{4k_n} \;, \qquad \text{for } n<\kappa_0 E p_+ \ell_s^2
\end{equation}
For example, for $p_+ > 2/(\kappa_0 E \ell_s^2)$, we get
\begin{equation} \label{eq:bound 2}
    \frac{1}{2k_1 |\alpha_{1,+}|} < \frac{4 k_1}{E|\Theta'(b)|} \;, \qquad \frac{1}{2k_2 |\alpha_{2,+}|} < \frac{4 k_2}{E|\Theta'(b)|} \;.
\end{equation}
Hence, we can combine \eqref{eq:bound1} and \eqref{eq:bound 2} to put an upper bound on the magnitude of \eqref{eq:oscillator product},
\begin{equation} \label{eq:bound G_osc}
    \begin{split}
        & |\tilde G^\Phi_{\rm osc}(x_i, x_f;p_+)| \leq 1 \qquad \qquad (\forall p_+) \;, \\
        &|\tilde G^\Phi_{\rm osc}(x_i, x_f;p_+)| \leq  \frac{1}{2k_1 |\alpha_{1,+}|} \cdot \frac{1}{2k_2 |\alpha_{2,+}|} < \frac{32}{E^2 |\Theta'(b)|^2 \ell_s^4 p_+^2 } \qquad \text{for }p_+ > \frac{2}{\kappa_0 E \ell_s^2} \;.
    \end{split}
\end{equation}

\subsubsection{Approximations}
\paragraph{Early Times} Early time singularities, $\mathcal T/M \to -\infty$, correspond to geodesics with a very large impact parameter, $b \gg M$. In this limit, we can approximate the oscillator determinant \eqref{eq:oscillator product} as follows.

In the large impact parameter regime, we have $r_t^2 \approx b^2 \gg M^2$, and since $r \geq r_t$ we can approximate \eqref{eq:A Schw} as
\begin{equation}
    \dot r(x^-) \approx \pm E\sqrt{1-b^2/r^2} \implies r(x^-) \approx \sqrt{b^2+(E   x^-)^2} \;,
\end{equation}
where we translated the affine parameter so $r(x^-=0)=r_t$. Hence, we get the pp wave profile
\begin{equation}
    A(x^-) \approx \frac{3 M E^2}{b^3} \left( \frac{1}{[1+(E x^-/b)^2]^{5/2}} + O(M/b) \right) \;,
\end{equation}
whose width in $x^-$ scales like $\delta x^-=b/E$. 

First, consider the oscillator modes whose wavelength $k_n^{-1}$ is much larger than $\delta x^-$,
\begin{equation} \label{eq:p+ ineq}
    \frac{p_+ \ell_s^2}{n} \gg b/E \;.
\end{equation}
These modes detect $A(x^-)$ as a shockwave localized at $x^-=0$. So we can evaluate the coefficients $\alpha_{n,\pm}$ using the shockwave approximation \cite{Dodelson:2020lal}. In other words, we replace $A(x^-)$ in \eqref{eq:GY ws} with 
\begin{equation}
    \frac{4 M E}{b^2} \delta(x^-) \;,
\end{equation}
then we compute the Gelfand-Yaglom solution with the boundary conditions \eqref{eq:yi},
\begin{equation}
    \begin{split}
        & y_{\pm, n}(x^-<0) = \frac{1}{k_n} \sin \left( k_n(x^--x^-_i) \right)  \;, \\
        & y_{\pm,n}(x^->0) = \alpha_{n,\pm} \exp(ik_n(x^--x^-_i)) + \beta_{n,\pm} \exp(-i k_n(x^--x_i^-)) \;.
    \end{split}
\end{equation}
$\alpha_{n,\pm}$ is easily evaluated as
\begin{equation}
    2i k_n\alpha_{n,\pm} = 1 \mp  2i\frac{M E}{b^2 k_n} (1-e^{2ik_n x_i^-}) \to 1 \mp 2i \frac{M E}{b^2 k_n} = 1\mp 2i \frac{M E \ell_s^2 p_+}{b^2 n} \;,
\end{equation}
where we took $x_i^- \to -\infty (1-i \epsilon)$. 

The shockwave approximation \eqref{eq:p+ ineq} eventually breaks down for sufficiently large $n$, for high-frequency modes with $k_n  \gtrsim  E/b = (\delta x^-)^{-1}$. But since $A(x^-)$ has height $A_0 \sim M E^2/b^3$,
\begin{equation}
    \frac{\delta x^- \cdot A_0}{k_n} \sim \frac{M E}{b^2} \cdot \frac{p_+ \ell_s^2}{n} \lesssim \frac{M}{b} \ll 1 \;.
\end{equation}
This is precisely the regime of validity of the Born approximation, which gives
\begin{equation}
    y_{n,\pm}(x^-) \approx \frac{1}{k_n} \sin (k_n(x^--x^-_i)) \pm \int_{x_i^-}^{x^-} du \frac{1}{k_n^2} \sin(k_n(x^- - u)) \sin(k_n(u-x^-_i)) \cdot A(u) \;.
\end{equation}
Sending $x^-_i \to -\infty(1-i\epsilon)$, $x^-_f \to \infty(1-i\epsilon)$, we get
\begin{equation}
    2i k_n \alpha_{n,\pm} \approx 1 \pm \frac{1}{2i k_n} \int_{-\infty}^\infty dx^- A(x^-)+O(n^{-2}) \approx 1\mp 2i \frac{ME\ell_s^2 p_+}{b^2n}  \cdot\left[1 +O(M/b) \right]+ O(n^{-2}) \;.
\end{equation}
The $O(n^{-2})$ are subleading in $M/b$ and can be safely ignored.

Hence, to leading order in $M/b$, we can use the approximation
\begin{equation}
    2i k_n \alpha_{n,\pm} \approx 1\mp 2i \frac{M E \ell_s^2 p_+}{b^2 n}
\end{equation}
for all values of $n$ and $p_+$. Plugging this into \eqref{eq:oscillator product} gives
\begin{equation} \label{eq:G_osc_large b}
    \begin{split}
        \lim_{r_i,r_f \to \infty} \tilde G^\Phi_{\rm osc}(x_i,x_f;p_+) \approx \prod_{n=1}^\infty \left[1+ \left(\frac{2p_+ \ell_s^2 M E}{b^2n} \right)^2\right]^{-1} = \frac{2\pi p_+ \ell_s^2 M E/b^2}{\sinh(2 \pi p_+ \ell_s^2 ME/b^2)}
    \end{split}
\end{equation}
This matches the AdS answer \cite{Dodelson:2020lal}, where the product of $\Gamma$ functions collapse into $\sinh$ in the special case of four dimensions.

\paragraph{Photon Sphere} We can also obtain analytic approximations to the oscillator determinant at very late times $\mathcal T/M \to \infty$, which correspond to geodesics that orbit the photon sphere for a very long time. These are trajectories with impact parameter $b \to b_c$. 

In this regime, the string approaches from infinity, then reaches the photon sphere $r\approx 3M$, where it spends a very long time in orbit
\begin{equation} \label{eq:t_PS}
    \Delta x^-_{\rm PS} =\frac{\sqrt{3} M \mathcal L(b)}{E} \gg \frac{M}{E} \;,
\end{equation}
during which the tidal profile is nearly constant
\begin{equation} \label{eq:A0}
    A(x^-) \approx A_0\equiv \frac{E^2}{3 M^2} \;.
\end{equation}
$\mathcal L(b)$ is the large log in \eqref{eq:photon-sphere-asymptotics}. For photon sphere kinematics, we work to leading order in $\mathcal L(b)$, dropping the subleading terms in this large quantity. 

When $p_+$ is large,
\begin{equation}
    p_+ > \frac{\sqrt{3} M}{\ell_s^2 E} \;,
\end{equation}
some of the oscillator modes $y_{n,+}$ become unstable,
\begin{equation} \label{eq:n_unstable}
    n < \frac{p_+ \ell_s^2 E}{\sqrt{3} M} \;.
\end{equation}
Worldsheet particle production for these modes is exponential, and can be computed using WKB \cite{Horowitz:1990sr}, giving the late time solution
\begin{equation}
    y_{n,+}(x^-) \approx \frac{1}{k_n} \exp \left( K_n \Delta x^-_{\rm PS} \right) \cdot [\sin(k_n(x^--x^-_i - \Delta x^-_{\rm PS} + O(1)) + \cos (k_n(x^-+x_i^-)) ] + O(e^{-K_n \Delta x^-_{\rm PS}}) \;,
\end{equation}
where
\begin{equation}
    K_n \equiv \sqrt{A_0-k_n^2} \;.
\end{equation}
This gives
\begin{equation}
    2i k_{n}\alpha_{n,+} \approx \exp [ \sqrt{3} M \mathcal L(b)(K_n -i k_n)/E + O(1)] \;,
\end{equation}
where we indicated corrections that do not grow with $\mathcal L(b)$, which we drop from now on. Summing over all the unstable modes \eqref{eq:n_unstable} gives an exponentially suppressed oscillator determinant for large $p_+$ \cite{Dodelson:2020lal},
\begin{equation} \label{eq:G_osc late_time p_large}
    |\tilde G_{\rm osc}(p_+)| \lesssim \exp\left(-  \frac{\pi}{4\sqrt{3} }\frac{\ell_s^2 \mathcal L(b)}{M} E p_+ \right) \;.
\end{equation}

To compute the finite piece of the string propagator, we need to compute the integral for $p_+$ below this cutoff. In particular, consider the opposite regime where
\begin{equation}
    p_+ \ll \frac{M}{\ell_s^2 E} \;.
\end{equation}
Then, we get
\begin{equation}
    A_0 \ll k_n^2 \;.
\end{equation}
All the modes are stable (oscillatory), and it can be easily checked from the geodesic equation $\dot r(x^-) = \pm E_0\sqrt{1-b^2 f(r)/r^2}$ that
\begin{equation}
    |\dot A(x^-)|\ll k_n^3 \;, \quad \, \text{and}  \, \qquad |\ddot A(x^-)| \ll k_n^4 \;.
\end{equation}
Hence, the WKB approximation can be safely used throughout the entire evolution for these modes (including the entrance and exit into the long photon sphere plateau), where it gives the solution satisfying the initial boundary conditions \eqref{eq:yi}
\begin{equation}
    y_{n, \pm}(x^-) = \frac{1}{[(k_n^2 \mp A(x_i^-)) \cdot (k_n^2 \mp A(x^-))]^{1/4}} \cdot \sin \left[ \int_{x_i^-}^{x^-} du \sqrt{k_n^2 \mp A(u)} \right] \;.
\end{equation}
This gives the coefficient
\begin{equation}
    2i k_n \alpha_{n,\pm} = \exp \left( i \int_{-\infty}^\infty dx^- [\sqrt{k_n^2\mp A(x^-)} - k_n] \right) \simeq \exp \left( i \int_{-\infty}^\infty dx^-  \left[ \mp\frac{A(x^-)}{2 k_n}  - \frac{A(x^-)^2}{8 k_n^3}\right] \right) \;,
\end{equation}
where we used $A(x^-) \ll k_n^2$ to approximate the phase. Hence, we get
\begin{equation} \label{eq:G_osc_late_time_p_small_intermediate}
    -4 k_n^2 \alpha_{n,+} \alpha_{n,-} \simeq \exp \left( -\frac{i}{4 k_n^3} \int_{-\infty}^\infty dx^-  A(x^-)^2\right) \simeq \exp \left( -\frac{i}{4 k_n^3} A_0^2 \Delta x_{\rm PS}^-\right) \;.
\end{equation}
The oscillator determinant can then be evaluated in a straightforward way
\begin{equation} \label{eq:G_osc late_time p_small}
    \tilde G_{\rm osc}(p_+) \approx \exp \left[i \frac{\zeta(3) \ell_s^6 p_+^3 E^3 \mathcal L(b)}{12 \sqrt{3} M^3} \right]
\end{equation}
where we used $\sum_{n=1}^\infty n^{-3} = \zeta(3)$. Even though $p_+ \ell_s^2 E/M \ll 1$, the factor of $\mathcal L(b)$ gives a large phase, such that the exponential cannot be Taylor expanded.

\addcontentsline{toc}{section}{References}
\bibliographystyle{style}
\bibliography{references.bib}

@article{Dodelson:2024bulkcone,
    author = "Dodelson, Matthew and Iossa, Cristoforo and Karlsson, Robin and Lupsasca, Alexandru and Zhiboedov, Alexander",
    title = "{Black hole bulk-cone singularities}",
    eprint = "2310.15236",
    archivePrefix = "arXiv",
    primaryClass = "hep-th",
    reportNumber = "CERN-TH-2023-192",
    doi = "10.1007/JHEP07(2024)046",
    journal = "JHEP",
    volume = "07",
    pages = "046",
    year = "2024"
}

@article{Dodelson:2020lal,
    author = "Dodelson, Matthew and Ooguri, Hirosi",
    title = "{Singularities of thermal correlators at strong coupling}",
    eprint = "2010.09734",
    archivePrefix = "arXiv",
    primaryClass = "hep-th",
    doi = "10.1103/PhysRevD.103.066018",
    journal = "Phys. Rev. D",
    volume = "103",
    number = "6",
    pages = "066018",
    year = "2021"
}

@article{Gubser:1996de,
    author = "Gubser, S. S. and Klebanov, Igor R. and Peet, A. W.",
    title = "{Entropy and temperature of black 3-branes}",
    eprint = "hep-th/9602135",
    archivePrefix = "arXiv",
    reportNumber = "PUPT-1598",
    doi = "10.1103/PhysRevD.54.3915",
    journal = "Phys. Rev. D",
    volume = "54",
    pages = "3915--3919",
    year = "1996"
}

@article{Klebanov:1997kc,
    author = "Klebanov, Igor R.",
    title = "{World volume approach to absorption by nondilatonic branes}",
    eprint = "hep-th/9702076",
    archivePrefix = "arXiv",
    reportNumber = "PUPT-1682",
    doi = "10.1016/S0550-3213(97)00235-6",
    journal = "Nucl. Phys. B",
    volume = "496",
    pages = "231--242",
    year = "1997"
}

@article{Maldacena:1997ih,
    author = "Maldacena, Juan Martin and Strominger, Andrew",
    title = "{Universal low-energy dynamics for rotating black holes}",
    eprint = "hep-th/9702015",
    archivePrefix = "arXiv",
    reportNumber = "UCSBTH-97-02",
    doi = "10.1103/PhysRevD.56.4975",
    journal = "Phys. Rev. D",
    volume = "56",
    pages = "4975--4983",
    year = "1997"
}

@article{Maldacena:1997re,
    author = "Maldacena, Juan Martin",
    title = "{The Large $N$ limit of superconformal field theories and supergravity}",
    eprint = "hep-th/9711200",
    archivePrefix = "arXiv",
    reportNumber = "HUTP-97-A097, HUTP-98-A097",
    doi = "10.4310/ATMP.1998.v2.n2.a1",
    journal = "Adv. Theor. Math. Phys.",
    volume = "2",
    pages = "231--252",
    year = "1998"
}

@article{Dray:1984ha,
    author = "Dray, Tevian and 't Hooft, Gerard",
    title = "{The Gravitational Shock Wave of a Massless Particle}",
    reportNumber = "Print-84-0773 (UTRECHT)",
    doi = "10.1016/0550-3213(85)90525-5",
    journal = "Nucl. Phys. B",
    volume = "253",
    pages = "173--188",
    year = "1985"
}

@article{tHooft:1996rdg,
    author = "'t Hooft, Gerard",
    title = "{The Scattering matrix approach for the quantum black hole: An Overview}",
    eprint = "gr-qc/9607022",
    archivePrefix = "arXiv",
    reportNumber = "THU-96-26",
    doi = "10.1142/S0217751X96002145",
    journal = "Int. J. Mod. Phys. A",
    volume = "11",
    pages = "4623--4688",
    year = "1996"
}

@article{Polchinski:2015cea,
    author = "Polchinski, Joseph",
    title = "{Chaos in the black hole S-matrix}",
    eprint = "1505.08108",
    archivePrefix = "arXiv",
    primaryClass = "hep-th",
    month = "5",
    year = "2015"
}

@article{Hubeny:2006yu,
    author = "Hubeny, Veronika E and Liu, Hong and Rangamani, Mukund",
    title = "{Bulk-cone singularities {\&} signatures of horizon formation in AdS/CFT}",
    eprint = "hep-th/0610041",
    archivePrefix = "arXiv",
    reportNumber = "DCPT-06-29, MIT-CTP-3775",
    doi = "10.1088/1126-6708/2007/01/009",
    journal = "JHEP",
    volume = "01",
    pages = "009",
    year = "2007"
}

@article{Gary:2009ae,
    author = "Gary, Mirah and Giddings, Steven B. and Penedones, Joao",
    title = "{Local bulk S-matrix elements and CFT singularities}",
    eprint = "0903.4437",
    archivePrefix = "arXiv",
    primaryClass = "hep-th",
    reportNumber = "CERN-PH-TH-2009-035, NSF-KITP-09-35",
    doi = "10.1103/PhysRevD.80.085005",
    journal = "Phys. Rev. D",
    volume = "80",
    pages = "085005",
    year = "2009"
}

@article{Maldacena:2015iua,
    author = "Maldacena, Juan and Simmons-Duffin, David and Zhiboedov, Alexander",
    title = "{Looking for a bulk point}",
    eprint = "1509.03612",
    archivePrefix = "arXiv",
    primaryClass = "hep-th",
    doi = "10.1007/JHEP01(2017)013",
    journal = "JHEP",
    volume = "01",
    pages = "013",
    year = "2017"
}

@article{Dolan:2011,
    author = "Dolan, Sam R. and Ottewill, Adrian C.",
    title = "{Wave Propagation and Quasinormal Mode Excitation on Schwarzschild Spacetime}",
    eprint = "1106.4318",
    archivePrefix = "arXiv",
    primaryClass = "gr-qc",
    doi = "10.1103/PhysRevD.84.104002",
    journal = "Phys. Rev. D",
    volume = "84",
    pages = "104002",
    year = "2011"
}

@article{Casals:2016qyj,
    author = "Casals, Marc and Nolan, Brien C.",
    title = "{Global Hadamard form for the Green function in Schwarzschild spacetime}",
    eprint = "1606.03075",
    archivePrefix = "arXiv",
    primaryClass = "gr-qc",
    doi = "10.1103/PhysRevD.108.044033",
    journal = "Phys. Rev. D",
    volume = "108",
    number = "4",
    pages = "044033",
    year = "2023"
}

@article{Buss:2017vud,
    author = "Buss, Cl{\'a}udia and Casals, Marc",
    title = "{Quantum correlator outside a Schwarzschild black hole}",
    eprint = "1709.05990",
    archivePrefix = "arXiv",
    primaryClass = "hep-th",
    doi = "10.1016/j.physletb.2017.11.048",
    journal = "Phys. Lett. B",
    volume = "776",
    pages = "168--173",
    year = "2018"
}

@article{Dolan:2009nk,
    author = "Dolan, Sam R. and Ottewill, Adrian C.",
    editor = "Uranga, A. M.",
    title = "{On an Expansion Method for Black Hole Quasinormal Modes and Regge Poles}",
    eprint = "0908.0329",
    archivePrefix = "arXiv",
    primaryClass = "gr-qc",
    doi = "10.1088/0264-9381/26/22/225003",
    journal = "Class. Quant. Grav.",
    volume = "26",
    pages = "225003",
    year = "2009"
}

@article{Grozdanov:2026cut,
    author = "Grozdanov, Sa{\v{s}}o and Valach, Samuel and Vrbica, Mile",
    title = "{Bouncing geodesics, black hole singularities, and singularities of thermal correlators}",
    eprint = "2603.15598",
    archivePrefix = "arXiv",
    primaryClass = "hep-th",
    doi = "10.1007/JHEP08(2026)022",
    journal = "JHEP",
    volume = "08",
    pages = "022",
    year = "2026"
}

@article{Grozdanov:2026ktq,
    author = "Grozdanov, Sa{\v{s}}o and Movrin, Vita and Valach, Samuel",
    title = "{Bouncing Geodesics, Singularities, and the Cavity Thermal Product Formula in Asymptotically Flat and de Sitter Black Holes}",
    eprint = "2606.11297",
    archivePrefix = "arXiv",
    primaryClass = "hep-th",
    month = "6",
    year = "2026"
}

@article{Grozdanov:2026lnc,
    author = "Grozdanov, Sa{\v{s}}o and Movrin, Vita and Valach, Samuel",
    title = "{Quasinormal modes as exterior probes of black hole interiors in our Universe}",
    eprint = "2608.13643",
    archivePrefix = "arXiv",
    primaryClass = "hep-th",
    month = "8",
    year = "2026"
}

@inbook{Penrose1976,
	address = {Dordrecht},
	author = {Penrose, Roger},
	booktitle = {Differential Geometry and Relativity: A Volume in Honour of Andr{\'e} Lichnerowicz on His 60th Birthday},
	doi = {10.1007/978-94-010-1508-0_23},
	isbn = {978-94-010-1508-0},
	pages = {271--275},
	publisher = {Springer Netherlands},
	title = {Any Space-Time has a Plane Wave as a Limit},
	url = {https://doi.org/10.1007/978-94-010-1508-0_23},
	year = {1976}
}

@article{Blau:notes,
    author = "Blau, Matthias",
    title = "{Plane Waves and Penrose Limits}",
    eprint = "http://www.blau.itp.unibe.ch/lecturesPP.pdf",
    archivePrefix = "",
    doi = "",
    journal = "",
    volume = "",
    pages = "",
    year = ""
}

@article{Horowitz:1989bv,
    author = "Horowitz, Gary T. and Steif, Alan R.",
    title = "{Space-Time Singularities in String Theory}",
    reportNumber = "UCSB-TH-89-38",
    doi = "10.1103/PhysRevLett.64.260",
    journal = "Phys. Rev. Lett.",
    volume = "64",
    pages = "260",
    year = "1990"
}

@article{Horowitz:1990sr,
    author = "Horowitz, Gary T. and Steif, Alan R.",
    title = "{Strings in Strong Gravitational Fields}",
    reportNumber = "UCSB-TH-90-08",
    doi = "10.1103/PhysRevD.42.1950",
    journal = "Phys. Rev. D",
    volume = "42",
    pages = "1950--1959",
    year = "1990"
}

@article{Metsaev:2001bj,
    author = "Metsaev, R. R.",
    title = "{Type IIB Green-Schwarz superstring in plane wave Ramond-Ramond background}",
    eprint = "hep-th/0112044",
    archivePrefix = "arXiv",
    reportNumber = "FIAN-TD-01-17",
    doi = "10.1016/S0550-3213(02)00003-2",
    journal = "Nucl. Phys. B",
    volume = "625",
    pages = "70--96",
    year = "2002"
}

@article{Berenstein:2002jq,
    author = "Berenstein, David Eliecer and Maldacena, Juan Martin and Nastase, Horatiu Stefan",
    title = "{Strings in flat space and pp waves from N=4 superYang-Mills}",
    eprint = "hep-th/0202021",
    archivePrefix = "arXiv",
    doi = "10.1088/1126-6708/2002/04/013",
    journal = "JHEP",
    volume = "04",
    pages = "013",
    year = "2002"
}

@article{Press:1971wr,
    author = "Press, William H.",
    title = "{Long Wave Trains of Gravitational Waves from a Vibrating Black Hole}",
    reportNumber = "OAP-262",
    doi = "10.1086/180849",
    journal = "Astrophys. J. Lett.",
    volume = "170",
    pages = "L105--L108",
    year = "1971"
}

@article{goebel_1972, title={Comments on the vibrations of a black hole}, volume={172}, ISSN={0004-637X}, DOI={10.1086/180898}, number={3}, journal={The Astrophysical Journal}, publisher={American Astronomical Society}, author={Goebel, C.J.}, year={1972}, month={Mar}, pages={L95} }

@ARTICLE{schutz,
       author = {{Schutz}, B.~F. and {Will}, C.~M.},
        title = "{Black hole normal modes - A semianalytic approach}",
      journal = {ApJL},
         year = 1985,
        month = apr,
       volume = {291},
        pages = {L33-L36},
          doi = {10.1086/184453},
       adsurl = {https://ui.adsabs.harvard.edu/abs/1985ApJ...291L..33S}
}

@article{Iyer:1986np,
    author = "Iyer, Sai and Will, Clifford M.",
    title = "{Black Hole Normal Modes: A {WKB} Approach. 1. Foundations and Application of a Higher Order {WKB} Analysis of Potential Barrier Scattering}",
    reportNumber = "Print-86-1482 (WASH. U., ST. LOUIS)",
    doi = "10.1103/PhysRevD.35.3621",
    journal = "Phys. Rev. D",
    volume = "35",
    pages = "3621",
    year = "1987"
}

@article{Iyer:1986nq,
    author = "Iyer, Sai",
    title = "{Black Hole Normal Modes: a WKB Approach. 2. Schwarzschild Black Holes}",
    reportNumber = "Print-86-1483 (WASH. U., ST. LOUIS)",
    doi = "10.1103/PhysRevD.35.3632",
    journal = "Phys. Rev. D",
    volume = "35",
    pages = "3632",
    year = "1987"
}

@article{Cardoso:2008bp,
    author = "Cardoso, Vitor and Miranda, Alex S. and Berti, Emanuele and Witek, Helvi and Zanchin, Vilson T.",
    title = "{Geodesic stability, Lyapunov exponents and quasinormal modes}",
    eprint = "0812.1806",
    archivePrefix = "arXiv",
    primaryClass = "hep-th",
    doi = "10.1103/PhysRevD.79.064016",
    journal = "Phys. Rev. D",
    volume = "79",
    number = "6",
    pages = "064016",
    year = "2009"
}

@article{Hadar:2022xag,
    author = "Hadar, Shahar and Kapec, Daniel and Lupsasca, Alexandru and Strominger, Andrew",
    title = "{Holography of the photon ring}",
    eprint = "2205.05064",
    archivePrefix = "arXiv",
    primaryClass = "gr-qc",
    doi = "10.1088/1361-6382/ac8d43",
    journal = "Class. Quant. Grav.",
    volume = "39",
    number = "21",
    pages = "215001",
    year = "2022"
}

@article{Fransen:2023eqj,
    author = "Fransen, Kwinten",
    title = "{Quasinormal modes from Penrose limits}",
    eprint = "2301.06999",
    archivePrefix = "arXiv",
    primaryClass = "gr-qc",
    doi = "10.1088/1361-6382/acf26d",
    journal = "Class. Quant. Grav.",
    volume = "40",
    number = "20",
    pages = "205004",
    year = "2023"
}

@article{Kapec:2024lnr,
    author = "Kapec, Daniel and Sheta, Ahmed",
    title = "{pp-waves and the hidden symmetries of black hole quasinormal modes}",
    eprint = "2412.08551",
    archivePrefix = "arXiv",
    primaryClass = "hep-th",
    doi = "10.1088/1361-6382/adecda",
    journal = "Class. Quant. Grav.",
    volume = "42",
    number = "15",
    pages = "155002",
    year = "2025"
}

@article{Christensen:1977jc,
    author = "Christensen, S. M. and Fulling, S. A.",
    title = "{Trace Anomalies and the Hawking Effect}",
    doi = "10.1103/PhysRevD.15.2088",
    journal = "Phys. Rev. D",
    volume = "15",
    pages = "2088--2104",
    year = "1977"
}

@article{Candelas:1980zt,
    author = "Candelas, P.",
    title = "{Vacuum Polarization in Schwarzschild Space-Time}",
    doi = "10.1103/PhysRevD.21.2185",
    journal = "Phys. Rev. D",
    volume = "21",
    pages = "2185--2202",
    year = "1980"
}

@article{Caribe:2023fhr,
    author = "Carib{\'e}, Jo{\~a}o G. A. and Jonsson, Robert H. and Casals, Marc and Kempf, Achim and Mart{\'\i}n-Mart{\'\i}nez, Eduardo",
    title = "{Lensing of vacuum entanglement near Schwarzschild black holes}",
    eprint = "2303.01402",
    archivePrefix = "arXiv",
    primaryClass = "quant-ph",
    reportNumber = "NORDITA 2023-022",
    doi = "10.1103/PhysRevD.108.025016",
    journal = "Phys. Rev. D",
    volume = "108",
    number = "2",
    pages = "025016",
    year = "2023"
}

@article{Leaver:1986gd,
    author = "Leaver, Edward W.",
    title = "{Spectral decomposition of the perturbation response of the Schwarzschild geometry}",
    doi = "10.1103/PhysRevD.34.384",
    journal = "Phys. Rev. D",
    volume = "34",
    pages = "384--408",
    year = "1986"
}

@article{Horvathy:2007nu,
    author = "Horvathy, P. A.",
    title = "{The Maslov correction in the semiclassical Feynman integral}",
    eprint = "quant-ph/0702236",
    archivePrefix = "arXiv",
    doi = "10.2478/s11534-010-0055-3",
    journal = "Central Eur. J. Phys.",
    volume = "9",
    pages = "1--12",
    year = "2011"
}

@article{Cohen:1985sm,
    author = "Cohen, Andrew G. and Moore, Gregory W. and Nelson, Philip C. and Polchinski, Joseph",
    title = "{An Off-Shell Propagator for String Theory}",
    reportNumber = "HUTP-85/A058, UTTG-16-85",
    doi = "10.1016/0550-3213(86)90148-3",
    journal = "Nucl. Phys. B",
    volume = "267",
    pages = "143--157",
    year = "1986"
}

@article{Russo:2002qj,
    author = "Russo, Jorge G. and Tseytlin, Arkady A.",
    title = "{A Class of exact pp wave string models with interacting light cone gauge actions}",
    eprint = "hep-th/0208114",
    archivePrefix = "arXiv",
    doi = "10.1088/1126-6708/2002/09/035",
    journal = "JHEP",
    volume = "09",
    pages = "035",
    year = "2002"
}

@article{Craps:2008bv,
    author = "Craps, Ben and De Roo, Frederik and Evnin, Oleg",
    title = "{Can free strings propagate across plane wave singularities?}",
    eprint = "0812.2900",
    archivePrefix = "arXiv",
    primaryClass = "hep-th",
    doi = "10.1088/1126-6708/2009/03/105",
    journal = "JHEP",
    volume = "03",
    pages = "105",
    year = "2009"
}

@inproceedings{Penrose19801O,
  title={1 – On Schwarzschild Causality—A Problem for “Lorentz Covariant” General Relativity},
  author={Roger Penrose},
  year={1980},
  url={https://api.semanticscholar.org/CorpusID:124027045}
}

@article{Cameron:2020itp,
    author = "Cameron, Peter and Dunajski, Maciej",
    title = "{On Schwarzschild causality in higher dimensions}",
    eprint = "2004.00086",
    archivePrefix = "arXiv",
    primaryClass = "gr-qc",
    doi = "10.1088/1361-6382/abb4b8",
    journal = "Class. Quant. Grav.",
    volume = "37",
    number = "22",
    pages = "225002",
    year = "2020"
}

@article{Monteverdi:2024xyp,
    author = "Monteverdi, Alessandro and Winstanley, Elizabeth",
    title = "{Some Addition Theorems for Spin-Weighted Spherical Harmonics}",
    eprint = "2410.23201",
    archivePrefix = "arXiv",
    primaryClass = "math-ph",
    doi = "10.3390/universe10120461",
    journal = "Universe",
    volume = "10",
    number = "12",
    pages = "461",
    year = "2024"
}

@article{Chon:2003gx,
    author = "Chon, Gayoung and Challinor, Anthony and Prunet, Simon and Hivon, Eric and Szapudi, Istvan",
    title = "{Fast estimation of polarization power spectra using correlation functions}",
    eprint = "astro-ph/0303414",
    archivePrefix = "arXiv",
    doi = "10.1111/j.1365-2966.2004.07737.x",
    journal = "Mon. Not. Roy. Astron. Soc.",
    volume = "350",
    pages = "914",
    year = "2004"
}

@article{Thorne,
  title = {Multipole expansions of gravitational radiation},
  author = {Thorne, Kip S.},
  journal = {Rev. Mod. Phys.},
  volume = {52},
  issue = {2},
  pages = {299--339},
  numpages = {0},
  year = {1980},
  month = {Apr},
  publisher = {American Physical Society},
  doi = {10.1103/RevModPhys.52.299},
  url = {https://link.aps.org/doi/10.1103/RevModPhys.52.299}
}

@article{Berti:2009kk,
    author = "Berti, Emanuele and Cardoso, Vitor and Starinets, Andrei O.",
    title = "{Quasinormal modes of black holes and black branes}",
    eprint = "0905.2975",
    archivePrefix = "arXiv",
    primaryClass = "gr-qc",
    doi = "10.1088/0264-9381/26/16/163001",
    journal = "Class. Quant. Grav.",
    volume = "26",
    pages = "163001",
    year = "2009"
}

@article{Hatsuda:2021gtn,
    author = "Hatsuda, Yasuyuki and Kimura, Masashi",
    title = "{Spectral Problems for Quasinormal Modes of Black Holes}",
    eprint = "2111.15197",
    archivePrefix = "arXiv",
    primaryClass = "gr-qc",
    reportNumber = "RUP-21-22",
    doi = "10.3390/universe7120476",
    journal = "Universe",
    volume = "7",
    number = "12",
    pages = "476",
    year = "2021"
}

@article{Strominger:2014pwa,
    author = "Strominger, Andrew and Zhiboedov, Alexander",
    title = "{Gravitational Memory, BMS Supertranslations and Soft Theorems}",
    eprint = "1411.5745",
    archivePrefix = "arXiv",
    primaryClass = "hep-th",
    doi = "10.1007/JHEP01(2016)086",
    journal = "JHEP",
    volume = "01",
    pages = "086",
    year = "2016"
}

@article{Correia:2025enx,
    author = "Correia, Miguel and Gopalka, Tushar and Isabella, Giulia and Wolz, Anna M.",
    title = "{Analyticity of the Black Hole S-Matrix}",
    eprint = "2511.11794",
    archivePrefix = "arXiv",
    primaryClass = "hep-th",
    month = "11",
    year = "2025"
}

@article{Jorstad:2024yzm,
    author = "Jorstad, Eivind and Pasterski, Sabrina",
    title = "{A comment on boundary correlators: soft omissions and the massless S-matrix}",
    eprint = "2410.20296",
    archivePrefix = "arXiv",
    primaryClass = "hep-th",
    doi = "10.1088/1361-6382/ae4da3",
    journal = "Class. Quant. Grav.",
    volume = "43",
    number = "6",
    pages = "065013",
    year = "2026"
}

@article{Camanho:2014apa,
    author = "Camanho, Xian O. and Edelstein, Jose D. and Maldacena, Juan and Zhiboedov, Alexander",
    title = "{Causality Constraints on Corrections to the Graviton Three-Point Coupling}",
    eprint = "1407.5597",
    archivePrefix = "arXiv",
    primaryClass = "hep-th",
    doi = "10.1007/JHEP02(2016)020",
    journal = "JHEP",
    volume = "02",
    pages = "020",
    year = "2016"
}

@article{Hawking:2016sgy,
    author = "Hawking, Stephen W. and Perry, Malcolm J. and Strominger, Andrew",
    title = "{Superrotation Charge and Supertranslation Hair on Black Holes}",
    eprint = "1611.09175",
    archivePrefix = "arXiv",
    primaryClass = "hep-th",
    doi = "10.1007/JHEP05(2017)161",
    journal = "JHEP",
    volume = "05",
    pages = "161",
    year = "2017"
}

@article{Donnay:2018ckb,
    author = "Donnay, Laura and Giribet, Gaston and Gonz{\'a}lez, Hern{\'a}n A. and Puhm, Andrea",
    title = "{Black hole memory effect}",
    eprint = "1809.07266",
    archivePrefix = "arXiv",
    primaryClass = "hep-th",
    reportNumber = "CPHT-RR021.042018",
    doi = "10.1103/PhysRevD.98.124016",
    journal = "Phys. Rev. D",
    volume = "98",
    number = "12",
    pages = "124016",
    year = "2018"
}

@article{Gonzo:2022tjm,
    author = "Gonzo, Riccardo and McLoughlin, Tristan and Puhm, Andrea",
    title = "{Celestial holography on Kerr-Schild backgrounds}",
    eprint = "2207.13719",
    archivePrefix = "arXiv",
    primaryClass = "hep-th",
    reportNumber = "CPHT-RR046.062022, HU-EP-22/26, SAGEX-22-24-E, TCD 22-04",
    doi = "10.1007/JHEP10(2022)073",
    journal = "JHEP",
    volume = "10",
    pages = "073",
    year = "2022"
}

@article{Long:2026rpq,
    author = "Long, Jiang and Qu, Zhan-Jia and Xiao, Hong-Yang",
    title = "{Carrollian correlators in black hole perturbation theory}",
    eprint = "2603.03033",
    archivePrefix = "arXiv",
    primaryClass = "gr-qc",
    month = "3",
    year = "2026"
}

@article{Donnay:2022wvx,
    author = "Donnay, Laura and Fiorucci, Adrien and Herfray, Yannick and Ruzziconi, Romain",
    title = "{Bridging Carrollian and celestial holography}",
    eprint = "2212.12553",
    archivePrefix = "arXiv",
    primaryClass = "hep-th",
    doi = "10.1103/PhysRevD.107.126027",
    journal = "Phys. Rev. D",
    volume = "107",
    number = "12",
    pages = "126027",
    year = "2023"
}

@article{Alday:2024yyj,
    author = "Alday, Luis F. and Nocchi, Maria and Ruzziconi, Romain and Yelleshpur Srikant, Akshay",
    title = "{Carrollian amplitudes from holographic correlators}",
    eprint = "2406.19343",
    archivePrefix = "arXiv",
    primaryClass = "hep-th",
    doi = "10.1007/JHEP03(2025)158",
    journal = "JHEP",
    volume = "03",
    pages = "158",
    year = "2025"
}

@article{Ruzziconi:2026bix,
    author = "Ruzziconi, Romain",
    title = "{Carrollian physics and holography}",
    eprint = "2602.02644",
    archivePrefix = "arXiv",
    primaryClass = "hep-th",
    doi = "10.1016/j.physrep.2026.03.005",
    journal = "Phys. Rept.",
    volume = "1182",
    pages = "1--87",
    year = "2026"
}

@article{batra2026nonpromptnullraycures,
      title={A non-prompt null ray cures its own singularity}, 
      author={Gauri Batra},
      year={2026},
      eprint={2610.22478},
      archivePrefix={arXiv},
      primaryClass={hep-th},
      url={https://arxiv.org/abs/2610.22478}, 
}

@misc{AlmheiriDodelsonKomargodski2026,
  title = {Quantum Gravity on the Bulk Cone},
  author = {Almheiri, A. and Dodelson, M. and Komargodski, Z.},
  note   = {To appear},
  year   = {2026}
}

@article{Blau:2006ar,
    author = "Blau, Matthias and Frank, Denis and Weiss, Sebastian",
    title = "{Fermi coordinates and Penrose limits}",
    eprint = "hep-th/0603109",
    archivePrefix = "arXiv",
    doi = "10.1088/0264-9381/23/11/020",
    journal = "Class. Quant. Grav.",
    volume = "23",
    pages = "3993--4010",
    year = "2006"
}

@article{Blau:2007db,
    author = "Blau, Matthias and Weiss, Sebastian",
    title = "{Penrose limits versus string expansions}",
    eprint = "0710.3480",
    archivePrefix = "arXiv",
    primaryClass = "hep-th",
    doi = "10.1088/0264-9381/25/12/125014",
    journal = "Class. Quant. Grav.",
    volume = "25",
    pages = "125014",
    year = "2008"
}

@article{Amati:1987uf,
    author = "Amati, D. and Ciafaloni, M. and Veneziano, G.",
    title = "{Classical and Quantum Gravity Effects from Planckian Energy Superstring Collisions}",
    reportNumber = "CERN-TH-4886/87",
    doi = "10.1142/S0217751X88000710",
    journal = "Int. J. Mod. Phys. A",
    volume = "3",
    pages = "1615--1661",
    year = "1988"
}

@article{Amati:1990xe,
    author = "Amati, D. and Ciafaloni, M. and Veneziano, G.",
    title = "{Higher Order Gravitational Deflection and Soft Bremsstrahlung in Planckian Energy Superstring Collisions}",
    reportNumber = "CERN-TH-5636-90",
    doi = "10.1016/0550-3213(90)90375-N",
    journal = "Nucl. Phys. B",
    volume = "347",
    pages = "550--580",
    year = "1990"
}

@article{Seiberg:2000gc,
    author = "Seiberg, N. and Susskind, Leonard and Toumbas, N.",
    title = "{Space-time noncommutativity and causality}",
    eprint = "hep-th/0005015",
    archivePrefix = "arXiv",
    reportNumber = "IASSNS-HEP-00-34, SU-ITP-00-14",
    doi = "10.1088/1126-6708/2000/06/044",
    journal = "JHEP",
    volume = "06",
    pages = "044",
    year = "2000"
}

@article{Cano:2025mht,
    author = "Cano, Pablo A. and David, Marina and van der Velde, Guido",
    title = "{Eikonal quasinormal modes of highly-spinning black holes in higher-curvature gravity: A window into extremality}",
    eprint = "2509.08664",
    archivePrefix = "arXiv",
    primaryClass = "gr-qc",
    doi = "10.1103/lypz-xz3y",
    journal = "Phys. Rev. D",
    volume = "113",
    number = "10",
    pages = "104025",
    year = "2026"
}

\end{document}